\documentclass[sigconf,nonacm]{acmart}

\usepackage{xcolor}

\AtBeginDocument{%
  \providecommand\BibTeX{{%
    \normalfont B\kern-0.5em{\scshape i\kern-0.25em b}\kern-0.8em\TeX}}}

\newcommand\vldbdoi{XX.XX/XXX.XX}

\newcommand\vldbavailabilityurl{http://vldb.org/pvldb/format_vol14.html}

\usepackage{mathtools}
\usepackage{amsthm}
\usepackage{amsmath}
\usepackage{array,multirow}
\usepackage{enumitem}
\usepackage{tabularx}
\usepackage{subcaption}
\usepackage{float}
\usepackage{thmtools} 
\usepackage{thm-restate}
\usepackage{hyperref}
\usepackage{cleveref}
\usepackage{commath}
\usepackage{algorithm}
\usepackage[noend]{algpseudocode}

\def\algbackskip{\hskip-\ALG@thistlm}

\theoremstyle{definition}
\declaretheorem[name=Lemma,numberwithin=section]{lemma}

\declaretheorem[name=Theorem,numberwithin=section]{theorem}
\newtheorem{corollary}{Corollary}[theorem]
\algrenewcommand\algorithmicrequire{\textbf{Input:}}
\algrenewcommand\algorithmicensure{\textbf{Output:}}
\newlength\myindent
\acmConference[Philadelphia '22]{Philadelphia '22: ACM SIGMOD/PODS International Conference on Management of Data}{June 13 -- June 15, 2022}{Philadelphia, PA} 
\acmBooktitle{Philadelphia '22: ACM SIGMOD/PODS International Conference on Management of Data, June 13 -- June 15, 2022, Philadelphia, PA}

\begin{document}

\title{Approximating Regret Minimizing Sets: A Happiness Perspective}

\author{Phoomraphee Luenam}
\email{pluenam@connect.ust.hk}
\affiliation{%
  \institution{Hong Kong University of Science and Technology}
  \streetaddress{P.O. Box 1212}
  \state{Hong Kong}
  \postcode{43017-6221}
}

\author{Yau Pun Chen}
\email{ypchen@connect.ust.hk}
\affiliation{%
  \institution{Hong Kong University of Science and Technology}
  \country{Hong Kong}}

\author{Raymond Chi-Wing Wong}
\email{raywong@cse.ust.hk}
\affiliation{%
  \institution{Hong Kong University of Science and Technology}
  \city{Hong Kong}
}

\renewcommand{\shortauthors}{Luenam, et al.} 

\begin{abstract}
A Regret Minimizing Set (RMS) is a useful concept in which a smaller subset of a database is selected while mostly preserving the best scores along every possible utility function. In this paper, we study the $k$-Regret Minimizing Sets ($k$-RMS) and Average Regret Minimizing Sets (ARMS) problems. $k$-RMS selects $r$ records from a database such that the maximum regret ratio between the $k$-th best score in the database and the best score in the selected records for any possible utility function is minimized. Meanwhile, ARMS minimizes the average of this ratio within a distribution of utility functions. Particularly, we study approximation algorithms for $k$-RMS and ARMS from the perspective of approximating the happiness ratio, which is equivalent to one minus the regret ratio. 

In this paper, we show that the problem of approximating the happiness of a $k$-RMS within any finite factor is NP-Hard when the dimensionality of the database is unconstrained and extend the result to an inapproximability proof for the regret. We then provide approximation algorithms for approximating the happiness of ARMS with better approximation ratios and time complexities than known algorithms for approximating the regret. We further provide dataset reduction schemes which can be used to reduce the runtime of existing heuristic based algorithms, as well as to derive polynomial-time approximation schemes for $k$-RMS when dimensionality is fixed. Finally, we provide experimental validation.
\end{abstract}

\maketitle




\section{Introduction}
Selecting a small subset of elements to represent a database is a fundamental problem which is of practical value since a database is often far too large for a typical user to search in it entirely. Some applications are selecting which products to advertise on a website or which phones to put on display. Two major techniques used for this purpose are top-$k$ queries and skyline queries. 

In top-$k$ queries, a utility function is given by the user and the top-$k$ records which maximize the utility function are returned. This has the advantage of giving the user a number of items to choose from, which is especially important when the exact utility function of the user is unclear or only vaguely known. However, the weakness of this type of query is that a utility function close to the true utility function must be known in advance when there can often be a wide variety of possible utility functions \cite{Chaudhuri:1999:ETS:645925.671359}. 

In skyline queries, all records not dominated by other records are returned, where a record dominates another record if all its coordinates are not worse and at least one of those coordinates is strictly better than those of the second record. While having the advantage of being able to function without the specification of any utility function, the skyline query does not effectively reduce the size of high dimensional databases \cite{Skyline}. In some cases, the skyline query returns the entire database when no record is dominated.

To avoid these limitations, the Regret Minimizing Set (RMS) query \cite{Nanongkai2010} was proposed to simultaneously possess the strengths of both types of queries,
resulting in many recent studies about this query in the database
community \cite{Asudeh:2017:ECR:3035918.3035932,Xie2019,Xie2018, Qi:2018:KQU:3243648.3230634,KesslerFaulkner:2015:KRQ:2831360.2831364,Unified}. In RMS, a subset of $r$ elements is chosen from a database of $n$ points such that the maximum regret ratio of any possible utility function between the best element in the database and the best element in the selected subset is minimized. This preserves the benefit of the bounded size of the top-$k$ query while keeping the skyline query's advantage of not requiring an exact utility function.

As comparing the elements in the selected subset to the best element in the database is a very demanding criterion, a relaxed version of the problem was proposed in $k$-Regret Minimizing Set ($k$-RMS) queries. In $k$-RMS, the selected subset is chosen such that the maximum regret ratio between the best element in the subset and the $k^{th}$ best element in the database for any given utility function is minimized.

Another approach is Average Regret Minimizing Sets (ARMS), where instead of minimizing the maximum regret ratio, the objective is instead minimizing the average regret ratio across a distribution of (possibly nonlinear) utility functions. The motivation behind using the average rather than the maximum regret is that optimizing the maximum may unfairly prioritize the least satisfied utility functions, while optimizing the average more properly satisfies the majority of utility functions. Since the expected regret of the distribution can be approximated with a sample with high confidence as the sample grows larger, it has been proposed to instead minimize the regret on a sample of $N$ utility functions. 

More recently, multiple papers have also considered the happiness maximization version of RMS problems. \cite{being_happy:ICDE:2020} studied the min-size version of RMS, where the goal is to find the smallest set that provides a given level of happiness, defined as equivalent to 1 minus the regret. Another, \cite{Qiu2018}, studied the happiness maximization version of $k$-RMS and provided an optimization to the greedy algorithm based on the monotonicity of the minimum happiness function. \cite{Storandt2019}, which studied both ARMS and its happiness maximization variant, makes particular use of the properties of the average happiness ratio to provide a better approximation ratio for the happiness variant over the original ARMS. However, there has not yet been a study that systematically compares the theoretical properties of the regret minimization and happiness maximization.

In this paper, we study the approximation of the happiness ratio for $k$-RMS and ARMS which arguably has more natural theoretical properties than approximating the regret. In particular, we are able to resolve the approximability status of happiness approximation for $k$-RMS completely while the approximability of the regret is still an open problem for some settings. Even more so, we provide several happiness approximation algorithms with provable bounds for $k$-RMS/ARMS that do not admit bounds on regret. 

\begin{enumerate}
\item For $k$-RMS, we will show the happiness ratio is NP-Hard to approximate when $d$, the dimensionality, is treated as an input for any fixed $k$ through a reduction from the set cover problem. We also extend this result to show that the problem of approximating the regret within a finite ratio is NP-Hard when $k$ and $d$ are treated as inputs, partially answering an open question posed in \cite{Kumar2018}.

\item We propose multiplicative and additive dataset reduction schemes for $k$-RMS from which we derive polynomial-time approximation schemes when $d$ is fixed. Together with previous results, this completely resolves the hardness of approximating the happiness of $k$-RMS for any $k$ and $d$, including unfixed $d$. We experimentally show that dataset reduction schemes can be used to significantly reduce the running time of existing heuristic based solvers for $k$-RMS while not significantly worsening the minimum happiness ratio/maximum regret ratio. For the largest settings tested, the reduction scheme was able to reduce the runtime by up to 93\% (from 4.2 hours to 16.7 minutes) while keeping happiness within 90\% of the original. 

\item For ARMS, we provide a $1-\frac{1}{e}$-approximation algorithm for the happiness of a function sample of size $N$ with a time complexity of $O(drNn)$, an improvement from the previous $O(dNn^3)$-time algorithm originally proposed for regret with no constant approximation bound \cite{Zeighami:2016:MAR:2882903.2914831}. We experimentally show that our algorithm scales efficiently up to a dataset of 1,000,000 points. 

\item For the special case of ARMS on a 2 dimensional dataset where the utility functions considered are linear, we provide an exact algorithm running in $O(n^2)$, an improvement from the $O(n^4)$ algorithm proposed in \cite{Zeighami:2016:MAR:2882903.2914831}. We also provide an approximation version running in $O(\frac{n}{\epsilon} + n \log n)$ where $\epsilon$ is the desired additive approximation ratio.

\end{enumerate}

The rest of the paper proceeds as follows:
Section 2 gives an overview of selected relevant work. Section 3 defines the $k$-RMS problem and presents our hardness results on the approximability of both the happiness and regret of $k$-RMS. Section 4 introduces additive and multiplicative dataset reduction schemes and extends them to polynomial time approximation schemes for $k$-RMS. Section 5 presents ARMS, and presents our proposed approximation algorithms.  Section 6 provides experimental results for 1) performance improvements from applying the reduction schemes before running previously proposed heuristic based algorithms for 1-RMS and 2) our proposed approximation algorithm for ARMS. Section 7 is the conclusion and discusses potential future work.
    
\section{Related Work}

In this section, we discuss some related work relevant to the RMS problem. We follow the naming conventions as in \cite{Xie2019}.

\subsection{RMS Problems}

RMS can be regarded as a special case of $k$-RMS when $k = 1$. Following \cite{Xie2019}, we categorize RMS algorithms for general dimensionality $d$ into two main classes based on whether there exist theoretically guaranteed results: heuristic approaches, and theoretical approaches.

\paragraph{\textbf{Heuristic Approaches}} RMS algorithms  that rely on heuristics can be further categorized into two subcategories: Linear Programming (LP) based algorithm and geometric algorithms. LP-based algorithms include Greedy \cite{Nanongkai2010} and ImpGreedy \cite{Xie2018}. Greedy \cite{Nanongkai2010} initializes RMS to the point with the best first dimensional value and iteratively inserts points that realize the current maximum regret ratio (computed with LP) until some defined stopping conditions are satisfied. ImpGreedy \cite{Xie2018} improves the efficiency of Greedy \cite{Nanongkai2010} by pruning nonessential LP computations. As shown in \cite{Qiu2018}, the efficiency could be further improved by performing randomized sampling on the input dataset before the greedy algorithms are executed. The geometric methods GeoGreedy and StoredList are greedy algorithms proposed by \cite{Peng2014}. The difference between GeoGreedy \cite{Peng2014} and Greedy \cite{Nanongkai2010} is that, in each iteration, the computational of maximum regret ratio is done with computational geometry methods instead of LP. As described in \cite{Peng2014}, StoredList is a materialization of GeoGreedy that pre-computes a set of candidates to run GeoGreedy on. 

\paragraph{\textbf{Theoretical Approaches}} Theoretically guaranteed approaches for RMS include Cube \cite{Nanongkai2010}, $\varepsilon$-Kernel \cite{Agarwal2017,cao_et_al:LIPIcs:2017:7056}, Sphere \cite{Xie2018}, HittingSet \cite{Agarwal2017,Kumar2018} and DMM \cite{Asudeh:2017:ECR:3035918.3035932}. 
Cube \cite{Nanongkai2010} constructs a solution set by dividing the data space into hypercubes based on the first $d-1$ dimensions and selecting the point within each hypercube with the largest coordinate in the $d^{th}$ dimension.
HittingSet \cite{Agarwal2017} transforms the RMS problem into a hitting set problem and applies an approximation algorithm from \cite{algorithmDesign}.
DMM works similarly to HittingSet but instead formulates the problem as a matrix min-max problem.
$\varepsilon$-Kernel \cite{Kumar2018} computes an $\varepsilon$-kernel on the original dataset to use as the input to the hitting set formulation and is more efficient than HittingSet. Sphere \cite{Xie2018} selects a small set of representative utility functions and includes points with high utilities for those functions.

\subsection{k-RMS Problems for $k \geq 2$ }

$k$-RMS, proposed by \cite{Chester:2014:CKM:2732269.2732275}, is a generalization of the RMS problem which relaxes the definition of regret to be computed against the $k^{th}$ best element along a given utility function rather than the single best element. Analogously to RMS, the goal of the $k$-RMS query is to minimize the maximum $k$-regret ratio over all possible utility functions while selecting up to $r$ elements. As previously mentioned, RMS can be viewed as a special case of $k$-RMS when $k$=1. The motivation behind this relaxation is that a user would often still be "happy" with even the second or third best choice, which makes optimizing for regret against the single best choice less practically useful. An added benefit of this relaxation is that it allows the dataset to be represented more succinctly. 

As $k$-RMS is a relaxation of the stricter $k$-regret query problem, it is possible to apply RMS algorithms such as CUBE and SPHERE which have known upper bounds $O(r^{-1/(d-1)})$ and $O(r^{-2/(d-1)})$ respectively to achieve an upper bound on the maximum $k$-regret ratio \cite{Nanongkai2010,Xie2018} (Note that these papers use $k$ to denote the number of points selected rather than $k$). However, these upper bounds may lie far from the optimal $k$-regret, and so these algorithms would not qualify as approximation algorithms for the $k$-RMS problem.

\paragraph{\textbf{Approximation Algorithms}}
Approximation algorithms for the RMS problem such as those proposed in \cite{Asudeh:2017:ECR:3035918.3035932} cannot in general be applied to $k$-RMS with the same approximation ratios, since it is possible that the relaxation may reduce the maximum regret ratio of the best possible solution. 

More recently, \cite{Agarwal2017} proposed a bicriteria approximation algorithm based on hitting sets for which the user can freely select the approximation ratio for the $k$-regret (requiring much larger run times for smaller approximation ratios) but may return more than the requested number of elements $r$ by up to a logarithmic factor. This deviates from the traditional setting of an approximation algorithm where generally only the optimized objective is allowed to differ from the optimal value by some factor.

Indeed, we show that it is impossible for there to be such an approximation algorithm for the general $k$-regret problem in arbitrary dimension unless P=NP.

\subsection{Average Regret Minimizing Sets}
ARMS was first studied in \cite{Zeighami:2016:MAR:2882903.2914831}, which defined the ARMS problem. ARMS was introduced to address issues with the original RMS problem. Specifically, RMS has the tendency to prioritize the least satisfied utility functions, which is often not representative of the majority of utility function. ARMS addresses this by instead minimizing the average regret ratio within a given distribution of utility functions. We further note that while the RMS problem was originally formulated using the set of linear utility functions, the ARMS is defined more generally for arbitrary utility functions.

\cite{Zeighami:2016:MAR:2882903.2914831} first showed that the regret could be closely approximated by its value on a sufficiently large sample of $N$ points and then provided an approximation algorithm running in $O(dNn^3)$ time based on the supermodularity and monotonicity of the average regret ratio function. They also provide exact algorithms for two dimensional datasets.

\cite{DBLP:conf/IEEEwisa/QiuZ18} further exploits the monotonicity of the average regret ratio to optimize the existing greedy algorithm with lazy evaluations. While this results in speedups, it does not improve the time complexity since there are worst case constructions that result in the same $O(dNn^3)$ runtime.

\cite{Storandt2019} studied the ARMS and its happiness variant, in the case of the space of linear functions. They provided a greedy algorithm with a $1-\frac{1}{e}$ approximation factor, whereas the existing result for ARMS only established an unbounded approximation bound dependent on the steepness of the average regret ratio. However, this requires computing volumes in $d$ dimensional spaces to do exactly, which may take up to $O(n^{d^2/4})$ time.

\section{$k$-RMS and Hardness of Approximation}

The goal of $k$-Regret Minimizing Sets ($k$-RMS) is to produce a small subset of a larger dataset that minimizes the maximum regret ratio (to be defined shortly). This results in a small representative set that ensures the highest regret is still within some acceptable ratio regardless of which utility function the user has. From the happiness perspective, this is equivalent to maximizing the lowest happiness ratio (defined to be 1 minus the regret ratio).

 In this section, we define $k$-RMS and present our hardness results on approximability. Specifically, we show that the happiness ratio of a $k$-RMS is inapproximability for any fixed $k$. This proof can be slightly changed to show the inapproximability of the regret of $k$-RMS (albeit treating $k$ as a parameter), partially resolving an open problem posed in \cite{Kumar2018}.
 
\newtheorem{defi}{Definition}

\subsection{Problem Definition}
\begin{table}[ht]
\caption{Notation Used in This Paper}
\centering
\begin{tabular}{ |p{1.5cm}|p{6cm}|  }
\hline
Symbol & Definition \\
\hline
\emph{D} & An input dataset \\
\emph{n} & \abs{\emph{D}}, the number of input points \\
\emph{d} & The number of dimensions \\
$\emph{p}_\emph{i}$ & The $\emph{i}^{th}$ point in \emph{D}\\
$\emph{p}_{\emph{i}}^{\emph{j}}$ & The $\emph{j}^{th}$ coordinate of $\emph{p}_{\emph{i}}$\\
\emph{R} & A subset of \emph{D}\\
\emph{r} & \abs{\emph{R}}, the number of elements in \emph{R} \\
\textbf{w} & A user weight vector\\
$\textbf{w}_{\emph{i}}$ & The $\emph{i}^{th}$ weight vector in a set \\
$\textbf{w}^j$ & The value in the $j^{th}$ dimension of \textbf{w} \\
$\emph{D}^{(\emph{k},\textbf{w})}$ & The $\emph{k}^{th}$ ranked point in \emph{D} with respect to \textbf{w}\\ 
$\emph{R}^{(\emph{k},\textbf{w})}$ & The $\emph{k}^{th}$ ranked point in \emph{R} with respect to \textbf{w}\\ 
\hline
\end{tabular}
\label{Tab:Tcr}
\end{table}
\newcommand{\w}[0]{\textbf{w}}
\newcommand{\p}[0]{\textbf{p}}

\DeclarePairedDelimiterX{\infdivx}[2]{(}{)}{%
  #1\;\delimsize\|\;#2%
}
\newcommand{\infdiv}{D\infdivx}

\vspace{\baselineskip}
Let $D$ be a database containing $n$ points in a $d$-dimensional space. All values of coordinates in the space are normalized such that the coordinates are real values in the range $[0,1]$, with at least one coordinate in each dimension being 1. A user weight vector or utility function is denoted by \w. We define the score of a point $p$ with respect to \w, denoted by $score(p,\w)$ as $p\cdot\w$ or, equivalently, $\sum_{j=1}^{d} p^j\w^j$. For simplicity, we assume without loss of generality, that $\w$ is normalized such that $\norm\w_1=1$ since it does not affect the problem, as both the numerator and denominator would be scaled by the same amount resulting in the same regret ratio. Now, we define $D^{(k,\w)}$ as the point in $D$ associated with the score with the $k^{th}$ rank in a sorted list of points' scores with respect to $\w$. Let the set of possible weight vectors be $\textbf{W}=\{\w \in [0,1]^d, \norm \w_1 =1\}$.

For a subset $R$ of $D$, the $k$-regret ratio with respect to a particular weight vector $\w$ is defined to be

\begin{defi}
$kreg(R,\w)= \max\left\{0,1-\frac{score(R^{(1,\w)},\w)}{score(D^{(k,\w)},\w)}\right\}$
\end{defi}

That is, if the best ranked point in $R$ is not worse than the $k^{th}$ ranked point in $D$ for a given $\w$, the $k$-regret ratio is 0. Otherwise, it is 1 minus the score of the best ranked point in $R$ divided by the score of the $k^{th}$ ranked point in $D$ with respect to \w. We define the $k$-regret ratio of a subset $R$ to be 

\begin{defi}
$kreg(R)=\max_{\w \in \textbf{W}} kreg(R,\w)$
\end{defi}
In other words, $kreg(R)$ is defined to be the maximum $k$-regret ratio with respect to any possible weight vector.

In a $k$-RMS query, the inputs are $D$, the database, and a positive integer $r$, the size of the returned $k$-RMS. A $k$-RMS is defined as the subset $R$ with size $r$ of $D$ such that $kreg(R)$ is minimum among all subsets of size $r$.

Analogously to the regret ratio functions, we define the happiness ratio functions as

\begin{defi}
$khapp(R,\w)= \min\left\{1,\frac{score(R^{(1,\w)},\w)}{score(D^{(k,\w)},\w)}\right\}$
\end{defi}

\begin{defi}
$khapp(R)=\min_{\w \in \textbf{W}} khapp(R,\w)$
\end{defi}

It is straightforward to verify that $khapp(R,\w) = 1 - kreg(R, \w)$ and $khapp(R) = 1 - kreg(R)$, implying that a $k$-RMS will also maximize happiness. Thus, the objective of happiness maximization is equivalent to regret minimization. However, it is possible to prove stronger theoretical results on the approximability of the happiness ratio, with the happiness being inapproximable even for $k=1$, which may be of theoretical interest. Furthermore, as we show in Section 4, the happiness maximization form of the problem admits multiplicative polynomial time approximation schemes for any fixed $d$.

\paragraph{Example} We have a dataset of 4 hotels as shown in Table 2. A 1-RMS of size 2 is $R=(A,C)$, achieving $kreg(R) = 1-\frac{0.5*0.8+0.5*0.35}{0.5*0.6+0.5*0.6} = 0.042 $. Here, the least happy utility function which determines the regret ratio is $\w = (0.5,0.5)$. 

\begin{table}[ht]
\caption{Example Dataset of Hotels and Scores}
\centering
\begin{tabular}{|l|l|l|}
\hline
Hotel & Stars & Price \\
\hline
A     & 0.8   & 0.35   \\
B     & 0.6   & 0.6   \\
C     & 0.35   & 0.8   \\
D     & 0.5   & 0.3   \\
\hline
\end{tabular}
\end{table}

\subsection{NP-Hardness of Approximating the Happiness of a $k$-RMS}

In this subsection, we prove that approximating the optimal $k$-happiness ratio of a $k$-RMS within any finite multiplicative ratio is NP-Hard even if $k$ is fixed. We begin by showing this for the special case $k=1$ through a reduction from the set cover problem, which is known to be NP-Hard. This result can then be extended to any larger value of $k$.

\begin{theorem}
Approximating the optimal $k$-happiness ratio of a $k$-RMS within any finite multiplicative ratio is NP-Hard for $k=1$ when treating $d$ as a parameter.
\end{theorem}
\begin{proof}

First, we define the set cover problem. For a set of items $U$ and a set of sets $T$ such that $\forall T_i \in T, T_i \subset U$ and a positive integer $r_{SC}$, does there exist a subset $S$ of $T$ with size no greater than $r_{SC}$ such that $\bigcup_{S_i \in S}S_i=U$. Let an instance of the set cover problem be denoted $I_{SC}(U,T)$.

We note that cases where there exists a member $U_j \in U$ such that $U_j \notin T_i \;\forall T_i \in T$ can be answered in $O(|U|+|T||U|)$ time by simply checking all the elements in each member of $T$. Since there is no set that covers $U_j$, it can be immediately concluded that the answer to such an instance is no.

For instances which do not fall into the previous category, from the given instance of the set cover problem, we will construct an instance of the $1$-RMS problem, $I_{1-RMS}$, and show that the existence of a polynomial-time approximation algorithm with a finite approximation factor would imply P=NP. Let the optimal $k$-happiness ratio for an instance $I_{1-RMS}$ be denoted $khapp(I_{1-RMS})$. Since a $k$-RMS maximizes the $k$-happiness ratio, the optimal value of $khapp(I_{1-RMS})$ is the greatest possible $k$-happiness ratio for $I_{1-RMS}(D,r)$.

 For $r$, the number of points to be selected in the $1$-RMS, set $r$=$r_{SC}$. Also, we construct $D$. Let $d$, the dimensionality of $D$, be equal to $|U|$. For each set $T_i \in T$, we construct a point $p$ in $D$, such that $p^j=1$ if $U_j \in T_i$ and $p^j=0$ otherwise. Let the points constructed from $T$ be known as the data points. Also, for each item $U_j \in U$, we construct a point $a_j$ where $a_j$ is the point such that $a_j^j=1$ and $a_j^l=0$ when $l \neq j$. Let the set of points constructed from $U$ be known as the axis points. From the construction, there are $|T|=k-1$ data points and $|U|$ axis points. Thus, $n$, the size of the database will be equal to $|T|+|U|=O(|T|+|U|)$. This takes $O(n\cdot d)=O(|T||U|+|U|^2)$ time.

We now make use of the following lemma. 
\begin{restatable}{lemma}{hmscase}
\label{thm:hmscase}
If the answer to $I_{SC}$ is no, $khapp(I_{1-RMS})=0$. Otherwise, the answer to $I_{SC}$ is yes and $khapp(I_{1-RMS})>0$.
\end{restatable}

\begin{proof}

Consider the case when there exists a subset $S$ of size no greater than $r_{SC}$ which contains all items in $U$. Then for each $S_i\in S$, we may select the point $p$ that was constructed from  $S_i$. Since by definition of being a solution to this instance of the set cover problem, we know that $|S|\leq r_{SC}$ and we would thus select not more than $r_{SC}=r$ points so it would be an acceptable selection for $R$ in the 1-RMS problem. We now consider the 1-happiness ratio of this selection. By definition of $S$ being a solution to $I_{SC}$, we know that $\forall U_j \in U, \exists S_i \in S$ such that $U_j \in S_i$. Thus, $\forall j \in \{1,...,d\}$, $\exists R_i \in R$ such that $R_i^j=1$ based on the construction of $D$.  Consider the 1-happiness ratio of $R$ with respect to a given weight vector $\mathbf{w}$. For any $\mathbf{w}$, we would necessarily have $(\exists R_i \in R, R_i^j=1) \implies (\exists R_i \in R, \mathbf{w} \cdot R_i >0) $ since any $\mathbf{w}$ has a positive coordinate in at least 1 dimension. Thus, $happ(\emph{R},\textbf{w})=\min\Big\{1,\frac{R^{(1,w)}\cdot\mathbf{w}}{D^{(1,w)}\cdot\mathbf{w}}\Big\}>0$. 

Consider the opposite case when there does not exist such a subset $S$. Assume there is some selection of $R$ of size $r$ with a 1-happiness ratio more than 0. Consider the weight vector $\mathbf{w}$ that points to an axis point $a_j$ with $||\mathbf{w}||_1=1$. Clearly, for any such weight vector, $D^{(1,\mathbf{w})}$ is at least the score of the axis point that maximizes, and thus, $D^{(1,\mathbf{w})}\cdot \mathbf{w} =1$. Now consider $R^{(1,\mathbf{w})}\cdot \mathbf{w}$. As it points to an axis point, $\mathbf{w}$ has the value 1 only in a single dimension and 0 in all others. Based on the construction of $D$, any coordinate of any point $R_i \in R$ must be either 1 or 0. Since we assumed $R$ has a 1-happiness ratio of more than 0, we must have $R^{(1,\mathbf{w})} \cdot \mathbf{w}=1$ as $R^{(1,\mathbf{w})} \cdot \mathbf{w}=0$ would contradict our assumption. This would imply that $\exists R_i \in R$ such that $R^j=1$. Since the 1-happiness ratio is no greater than the minimum across all $d$ weight vectors that point to an axis point, we must have that $\forall j \in \{1,...,d\}, \exists R_i\in R$ such that $R_i^j=1$. However, then we could select each set $U_i$ that corresponds to a point $R_i$ in $R$ during the construction of $D$ for $I_{SC}$ as $S$ (We have excluded cases where there is an item in $U$ that is covered by no set in $T$ so if an axis point was chosen in $R$, it may be replaced with another data point which either dominates or is equivalent to the axis point). This would contradict the assumption that $S$ does not exist. Therefore, there is no selection of $R$ with 1-happiness ratio more than 0. Thus, in this case, $khapp(I_{1-RMS})=0$. 
\end{proof}

Applying the lemma, any polynomial-time approximation algorithm with a finite multiplicative positive approximation ratio for the 1-happiness ratio would be able to distinguish the two cases of whether or not there exists a set cover, and its existence would imply P=NP. Thus, the problem of approximating the 1-happiness ratio of the optimal 1-RMS to a finite factor is NP-hard.
\end{proof}

\begin{corollary}
Approximating the $k$-happiness ratio of a $k$-RMS within any finite multiplicative ratio is NP-Hard for any fixed $k$ when treating $d$ as a parameter.
\end{corollary}
\begin{proof}
Since we can approximate the 1-happiness of 1-RMS with an approximation for the $k$-happiness of a $k$-RMS by making $k$ copies of every point, approximating the $k$-happiness of a $k$-RMS must also be NP-Hard for any other $k$.
\end{proof}

The proof of Theorem 3.1 can be changed slightly to show the NP-Hardness of approximating the regret of $k$-RMS as well when $k$ is treated as a parameter, in contrast to the happiness ratio where the result applies to any fixed $k$. 

\begin{restatable}{theorem}{krms}
\label{thm:krms}
Approximating the regret of a $k$-RMS within any finite multiplicative ratio is NP-Hard when $k$ and $d$ treated as parameters.
\end{restatable}

\begin{proof}
The proof follows from the same reduction from the set cover problem as in Theorem 3.1, with the changes that $k$ is set to $|T|+1$ (where $|T|$ is the size of the set of sets) and that we construct $k$ copies of each axis point instead of only one. 
\end{proof}

\newcommand{\W}{\textbf{W}}

\section{Dataset Reduction Schemes and Polynomial Time Approximation Schemes} 
Having shown that both the happiness and regret of $k$-RMS are NP-Hard to approximate in the general case, we now introduce dataset reduction schemes to improve the runtime of existing heuristic based approaches. We extend these reduction schemes to show that polynomial time approximations algorithms for the happiness of $k$-RMS are achievable for fixed dimensionality $d$ that allow any desired multiplicative or additive approximation factor. 

While these approximation schemes are computationally infeasible, the dataset reduction schemes can be prior to existing heuristic based algorithms which often have poor scalabilty. In Section 6, we show experimental validation for the efficiency boost from applying the additive and multiplication reduction schemes on the performance of selected heuristic algorithms. 

\subsection{Dataset Reduction Schemes}
We begin by defining the dataset reduction schemes and proving several properties about them that are used in the polynomial time approximation schemes. 

A dataset reduction scheme is an algorithm that takes in a dataset $D$ as an input and outputs a new dataset $D'$ such that $|D'| \leq |D|$ and each point in $D'$ corresponds to an original point in $D$. We present two reductions schemes: the \textbf{Additive Reduction Scheme} and the \textbf{Multiplicative Reduction Scheme}. In this subsection, we will prove several useful properties about the reduction schemes that will ultimately be used in the proof of the polynomial time approximation schemes.

\subsubsection{\textbf{Additive Reduction Scheme}} Given an additive approximation factor $\varepsilon$, let $\varepsilon'= \frac{\varepsilon}{d}$. We create a new dataset $D'$ where the coordinates of each point are rounded down to nearest multiple of $\varepsilon'$. Formally, for each point $p$ in $D$, we create a point $p'$ in $D'$ such that each coordinate $p'^j$ of $p'$, we set $p'^j$ to the greatest multiple of $\varepsilon'$ no greater than $p^j$.

\subsubsection{\textbf{Multiplicative Reduction Scheme}}
    Given a multiplicative approximation factor $1-\varepsilon$, let $\varepsilon^*= \frac{\varepsilon}{2}$. We create a new dataset $D'$ where the coordinates of each point are rounded down to nearest power of $(1-\varepsilon^*)$ and if this value would be less than $\frac{\varepsilon^*}{d}$, we set it to 0. Formally, for each point $p$ in $D$, we create a point $p'$ in $D'$ such that each coordinate $p'^{j}$ of $p'$, we set $p'^j$ to the greatest power of $(1-\varepsilon^*)$ no greater than $p^j$ if this value is at least $\frac{\varepsilon^*}{d}$ and $p'^{j}$ is set to 0 otherwise.

We begin by proving bounds on the size of the reduced dataset from both schemes.

\begin{lemma}
The output dataset $D'$ from the \textbf{Additive Reduction Scheme} has size at most $(\frac{d}{\varepsilon}+1)^d$.
\end{lemma}
\begin{proof}
Since originally each coordinate ranged from 0 to 1, for each coordinate there can be at most $\frac{1}{\varepsilon'}+1$ distinct values for coordinates and since there are only $d$ dimensions, at most $(\frac{1}{\varepsilon'}+1)^d=(\frac{d}{\varepsilon}+1)^d$ distinct points can exist in $D'$. 
\end{proof}

\begin{lemma}
The output dataset $D'$ from the \textbf{Multiplicative Reduction Scheme} has size at most $(\log_{1-\varepsilon/2} \frac{\varepsilon}{2d}+2)^d$.
\end{lemma}
\begin{proof}
There are at most $\log_{1-\varepsilon^*} \frac{\varepsilon^*}{d}+2$ distinct values of coordinates since for powers of $(1-\varepsilon^*)$ we have $(1-\varepsilon^*)^k < \frac{\varepsilon^*}{d}$ for any $k > \log_{1-\varepsilon^*} \frac{\varepsilon^*}{d}$ and the only other allowed values are 1 and 0. Thus, there are at most $(\log_{1-\varepsilon^*} \frac{\varepsilon^*}{d}+2)^d=(\log_{1-\varepsilon/2} \frac{\varepsilon}{2d}+2)^d$ distinct coordinate tuples in $D'$.
\end{proof}

Next, we prove that the happiness ratio of the optimal set in the reduced dataset is at worst some (additive or multiplicative) factor worse than the happiness ratio of the optimal set in the original dataset.

\begin{lemma}
For the \textbf{Additive Reduction Scheme}, the optimal $k$-regret minimizing set $O$ in $D$ corresponds to a subset $O'$ in $D'$ such that the $k$-happiness ratio of $O'$ is at worse that of $R$ minus $\varepsilon$.
\end{lemma}
\begin{proof}
Consider an optimal $k$-regret minimizing set $O$ for any instance of the problem. Note that $O$ also optimizes $k$-happiness. Let $O'$ be the corresponding subset in $D'$. We must have that $O'^j_i \geq O^j_i -\varepsilon'$ for any point because of the way we do the rounding. Thus, for any weight vector $\w$, we will have that $\w \cdot O'_i \geq \sum_{j=1}^d \w^j O'^j_i \geq \sum_{j=1}^d w^j (O_i^j-\varepsilon')=\w\cdot O_i - \varepsilon' \sum_{j=1}^d \w^j \geq \w\cdot O_i -\varepsilon' $. Then it follows that $\max_{O'_i \in O'} \w\cdot O'_i \geq \max_{O_i \in O} \w\cdot O_i- \varepsilon'$ $\implies$ $\frac{\max_{O'_i \in O'} \w\cdot O'_i}{\max_{p \in D} \w\cdot p} \geq \frac{\max_{O_i \in O} \w\cdot O_i}{\max_{p \in D} \w\cdot p}-\frac{ \varepsilon'}{\max_{p \in D} \w\cdot p}\geq \frac{\max_{O_i \in O} \w\cdot O_i}{\max_{p \in D} \w\cdot p}- \varepsilon'd =\frac{\max_{O_i \in O} \w\cdot O_i}{\max_{p \in D} \w\cdot p}- \varepsilon$, and thus the happiness ratio of $O'$ is at worst that of $O$ minus $\varepsilon$. 
\end{proof}

For the corresponding proofs for the multiplicative version, we will make use of the following lemma.
\begin{lemma}
For $x\geq \frac{1}{d}$ and $\varepsilon>0$, $x-\frac{\varepsilon}{d}\geq (1-\varepsilon)x$
\end{lemma}
\begin{proof}
$\varepsilon x > \frac{\varepsilon}{d} \implies (1-\varepsilon)x = x-\varepsilon x \leq x - \frac{\varepsilon}{d}$
\end{proof}
\begin{lemma}
For the \textbf{Multiplicative Reduction Scheme}, if $|O|\geq d$, the optimal $k$-regret minimizing set $O$ in $D$ corresponds to a subset $O'$ in $D'$ such that the $k$-happiness ratio of $O'$ is at worse that of $O$ times $(1-\varepsilon)$.
\end{lemma}
\begin{proof}
Consider an optimal $k$-regret minimizing set $O$ for any instance of the problem. Again, let $O'$ be the corresponding subset in $D'$.

 We first note that since $|O|\geq d$, $O$ must perform at least as well as selecting $d$ points with 1 in each coordinate, and $||\w||_1=1$, $\max\limits_{O_i \in O} \w \cdot O_i \geq \frac{1}{d} $ which implies $ \max\limits_{O_i \in O} \w \cdot O_i-  \frac{\varepsilon^*}{d} \geq (\max\limits_{O_i \in O} \w \cdot O_i) (1-\varepsilon^*)$ by Lemma 4.4.
    
Now consider the effect of setting some coordinates to 0 due to their value being less than $\frac{\varepsilon^*}{d}$. The reduction in the value of $\w \cdot O'_i$ is no more than $\frac{\varepsilon^*}{d} \sum_{j=1}^d \w^j  \leq \frac{\varepsilon^*}{d} $ in the case we set all coordinates of $O'_i$ to 0. Then, since any coordinate is scaled down by at most $(1-\varepsilon^*)$, we must have 
 \[\max\limits_{O'_i \in O'} \w \cdot O'_i \geq (1-\varepsilon^*)( \max\limits_{O_i \in O} \w \cdot O_i -\frac{\varepsilon^*}{d})\]


Thus, \[\begin{split}\max\limits_{O'_i \in O'}\ w \cdot O'_i &\geq (1-\varepsilon^*)^2 \max\limits_{O_i \in O} w \cdot O_i =  (1-2\varepsilon^*+(\varepsilon^*)^2) \max\limits_{O_i \in O} \w \cdot O_i  \\ &\geq (1-2\varepsilon^*) \max\limits_{O_i \in O} \w \cdot O_i  = (1-\varepsilon) \max\limits_{O_i \in O} \w \cdot O_i  \end{split}\] 


Then it follows that $\max_{O'_i \in O'} \w\cdot O'_i \geq (\max_{O_i \in O} \w\cdot O_i)(1- \varepsilon)$ $\implies$ $\frac{\max_{O'_i \in O'} \w\cdot O'_i}{\max_{p \in D} \w\cdot p} \geq(1-\varepsilon)(\frac{\max_{O_i \in O} \w\cdot O_i}{\max_{p \in D} \w\cdot p})$, and thus the $k$-happiness ratio of $O'$ is at least that of $O$ times $(1-\varepsilon)$. 
\end{proof}

Finally, we prove that a set in the reduced dataset corresponds to some set in the original with at least the same happiness ratio for either reduction scheme.

\begin{lemma}
A set $A'$ in $D'$ corresponds to a set $A$ in $D$ with at least the same happiness ratio.
\end{lemma}
\begin{proof}
Each point $p'$ in $A'$ corresponds to some point $p(p')$ in $D$ such that $p$ (weakly) dominates $p'$. Taking $A=\bigcup_{p' \in A'} \{p(p')\}$, it follows that the happiness ratio of $A$ is at worst the happiness ratio of $A'$. 
\end{proof}

\subsection{Polynomial Time Approximation Schemes}
\label{subsec:polynomialTimeApproximation}

In this subsection, we present our polynomial-time approximation schemes for approximating the happiness of $k$-RMS. While these approximation schemes are computationally infeasible, they nevertheless demonstrate that when $d$ is fixed, approximating the $k$-happiness can be done in polynomial time for any desired approximation ratio. This result resolves the multiplicative approximability status of $k$-happiness for all $k$ and $d$ - it is NP-Hard to multiplicatively approximate to any constant ratio when $d$ is unfixed, and can be approximated in polynomial time to any desired ratio when $d$ is fixed.

We make use of the following result which trivially follows from Theorem 3.2 of \cite{Agarwal2017}.

\begin{lemma}
For any dataset $D$ with $n$ points, a subset $R$ of size $O(\frac{1}{\varepsilon^{(d-1)/2}})$ with happiness ratio at least $1-\varepsilon$ can be computed in $O(n+\frac{1}{\varepsilon^{d-1}})$.
\end{lemma}
    
\textbf{Polynomial-Time Approximation Scheme}
\label{subsec:additivePolyTimeApproScheme}

\begin{enumerate}
    \item We first attempt applying Lemma 4.7 to compute a subset $S$ of $D$ with happiness ratio at least $1-\frac{1}{\varepsilon^{(d-1)/2}}$. If $r$ is at least the resulting size, we can immediately return $S$ with additive approximation factor $\varepsilon$ and multiplicative factor $1-\varepsilon$ . Otherwise, $r$ is of $O(\frac{1}{\varepsilon^{(d-1)/2}})$ size, so $r\leq \frac{c}{\varepsilon^{(d-1)/2}}$ for some constant $c$.
    
    \item \textbf{Reduction Scheme} Given the approximation factor $\varepsilon$, apply the additive or multiplicative reduction scheme on $D$ to get the reduced dataset $D'$.
    
    \item Consider any possible combination of $r$ points in $D'$ and compute the happiness ratio for each combination, applying algorithm from Lemma 2 in \cite{Agarwal2017} which runs in $O(n^{2d-1})$ time. Choose the combination that results in the best happiness.
    
    \item Construct the set $A$ using Lemma 4.6 and return it.
\end{enumerate}

\begin{theorem}
The \textbf{Polynomial-Time Approximation Scheme} runs in polynomial time and results in an additive $\varepsilon$ factor happiness approximation with the additive reduction scheme or a multiplicative $1-\varepsilon$ factor happiness approximation with the multiplicative reduction scheme.
\end{theorem}
\begin{proof}
Steps 1, 2, and 4 run in polynomial time. As for step 3, since $|D'|$ is constant given fixed $d$ and $\varepsilon$ ($(\frac{d^2}{\varepsilon}+1)^d$ for the additive scheme [Lemma 4.1] and $(\log_{1-\varepsilon'} \frac{\varepsilon'}{d^3}+2)^d$ for the multiplicative scheme [Lemma 4.2]), there are most ${|D|}^r \leq {|D'|}^{\frac{c}{\varepsilon^{(d-1)/2}}} $ combinations, which is polynomial in (and actually independent of) the input size for fixed $\varepsilon$ and $d$. Checking the happiness ratio of each will again take polynomial time so the total algorithm runs in polynomial time.

The correctness of the approximation scheme follows from the lemmas in the previous subsection. The happiness ratio of $A'$ is at worst an additive factor $\varepsilon$ worse than the optimal set $O$ in $D$ (Lemma 4.3) or multiplicative factor 1-$\varepsilon$ (Lemmas 4.5), and the returned set then has approximation ratio at worst $\varepsilon$ (Lemma 4.6).
\end{proof}

Note that the additive scheme can also be applied to approximate the regret of $k$-RMS with the same $\varepsilon$ approximation factor (the $k$-regret ratio of the returned set would be at worst $OPT+\varepsilon$). This follows from the result that $kreg(R)=1-khapp(R)$.

Unlike the additive scheme, the multiplicative scheme cannot be applied to the $k$-regret ratio with any finite bounds. A trivial example where this happens is a dataset of two points where one is slightly better in all coordinates, but both are rounded to the same point - the regret for one point is $0$, but some positive value for the other.

\section{Average Regret Minimizing Sets}
Besides $k$-RMS, an alternative approach to select a representative subset is Average Regret Minimizing Sets (ARMS). Instead of minimizing the maximum regret ratio, ARMS minimizes the average regret ratio, which may be more appropriate for situations where it is less necessary to optimize the regret of relatively rare utility functions.

In this section, we define the ARMS problem. We then provide an approximation algorithm with approximation ratio 1-$\frac{1}{e}$ for the average happiness ratio of a function sample for the general case. Finally, we provide an exact $O(n^2)$ algorithm for the special case of linear utilities in 2 dimensions, which can be turned into a $O(\frac{n}{\epsilon})$ additive $\epsilon$-approximation algorithm. 

\subsection{Problem Definition}
Similar to $k$-RMS as described in Section 3.1, in ARMS we are given a dataset $D$ of $n$ points in a $d$-dimensional space. The score of a point $p$ against a utility function \textbf{w} is again denoted $score(p,\w)$. However, in contrast to $k$-RMS, current formulations of the ARMS query \cite{Zeighami:2016:MAR:2882903.2914831,Qiu2018} also require the set of utility functions to consider, $F$, and its probability distribution as inputs. The utility functions considered are not necessarily linear. 

Formally, for an ARMS query, the inputs are $D$, the database, a positive integer $r$, the size of the returned ARMS, and the set of utility functions considered, $F$ along with its probability distribution $\Theta$. An ARMS is defined as the subset $R$ with size $r$ of $D$ such that the average regret ratio, $arr(R)$, is minimum among all subsets of size $r$. For simplicity, each function $f \in F$ is assumed to be computable in $O(d)$ time.

Here, given $F$ and $\eta$ (the probability density function corresponding the the probability distribution $\Theta$), $arr(R)$, the Average Regret Ratio, is defined as

\begin{defi}
$arr(R)= \int_{f \in F} reg(R,f) \eta(f) df $
\end{defi}

where the definition of the regret ratio, $reg(R,f)$, is generalized for possibly non-linear functions as

\begin{defi}
$reg(R,f)= \max\left\{0,1-\frac{\max_{p \in R} f(p)}{\max_{p \in D} f(p)}\right\}$
\end{defi}

In other words, the Average Regret Ratio, $arr(R)$, is the expectation of  $reg(R)$, $E[reg(R)]$ when considering the set of utility functions $F$. The Average Happiness Ratio, $ahr(R)$, is then simply $1-arr(R)$.

\paragraph{Example} Consider again the hotels shown in Table 2. Suppose the set of utility functions considered are the linear functions in Table 3. An ARMS of size 2 is $R=(A,C)$, achieving $arr(R) = 0.6\cdot0+0.2\cdot0+0.2\cdot(1-\frac{0.5*0.8+0.5*0.35}{0.5*0.6+0.5*0.6}) = 0.008 $. Here utility functions 1 and 2 are fully satisfied by hotels A and C respectively.

\begin{table}[ht]
\caption{Example Set of Utility Functions and Probabilities}
\centering
\begin{tabular}{|l|l|l|l|}
\hline
Utility Function & Stars Weight & Price Weight & Probability\\
\hline
1     & 0   & 1  &0.6 \\
2     & 1   & 0  &0.2  \\
3     & 0.5   & 0.5  &0.2  \\
\hline
\end{tabular}
\end{table}

\subsection{Approximating Average Happiness Ratio of a Sample of Functions }

While the definition previously given describes the general form of an ARMS problem, it leaves open the question of how to  encode the input, since the space of considered functions is unrestricted. In many practical cases, it is possible to instead compute an approximation, $arr^*(R)$, with arbitrary accuracy by computing the average happiness ratio for a random sampling $S$ of $N$ utility functions for sufficiently large $N$ (see \cite{Zeighami:2016:MAR:2882903.2914831} for precise guarantees and discussion):

\begin{defi}
\[arr^*(R,S) = \dfrac{1}{N}\sum_{f\in S} \left( 1-\dfrac{\max_{p\in R} f(p)}{\max_{p\in D} f(p)} \right)\]
\end{defi}

While an approximation algorithm was provided for $arr^*(R)$, in  \cite{Zeighami:2016:MAR:2882903.2914831}, the approximation ratio is given as $\frac{e^t-1}{t}$, where $t=\frac{s}{1-s}$ and $s$ is the "steepness" (maximum marginal decrease) of the Average Regret Ratio. This approximation ratio is not fixed and is potentially quite bad (indeed it is not defined for $s=1$). 
Meanwhile, for $ahr^*(R)$, defined analogously, there exists a probabilistic approximation algorithm with fixed approximation ratio $1-\frac{1}{e}$ as shown in \cite{Storandt2019} the case where $F$ is the infinite set of linear utility functions. This proof extends to general functions using a similar approach.

In \cite{Zeighami:2016:MAR:2882903.2914831}, it was shown that the sampled average regret ratio, $arr^*(\cdot,S)$ is monotonically decreasing and has the supermodularity property. Since $ahr^*(R,S) = 1- arr^*(R,S)$, it trivially follows that $ahr^*(\cdot , S)$ is submodular and monotonically increasing. 
It then follows from the result in \cite{Nemhauser1978} on the maximization of monotonically increasing submodular functions that the maximum value of $ahr^*(R,S)$ can be approximated to a $1-\frac{1}{e}$ ratio by iteratively greedily choosing $R_{s+1}=R_s \cup {x}$ where $x$ is the point that results in the greatest $ahr^*(R_s \cup {x}, S)$, starting from the empty set $R_0 = \varnothing $. This greedy approach will form the backbone for the approximation algorithms provided in this section.

In \cite{Zeighami:2016:MAR:2882903.2914831}, an approximation algorithm for minimizing $arr^*(R,S)$ running in $O(dNn^3)$ time was given with the aforementioned unbounded approximation ratio $\frac{e^t-1}{t}$. Because their algorithm is based on minimizing supermodular functions, it iteratively removes points rather than adding them. Simply changing the algorithm to iteratively add while maximizing $ahr^*(R,S)$ already results in the improved time complexity of $O(drNn^2)$ with the fixed multiplicative approximation ratio of $1-\frac{1}{e}$. However, here we give an algorithm with further improved runtime bounds. 

The main idea of the approximation algorithm is to keep track of the improvement in average happiness ratio gained by adding each point, $\Delta_x$, and iteratively adding the point $p_x$ with the highest $\Delta_x$. To do this, within each iteration, after adding each point $x$, we recalculate the happiness $\delta_{j,i}$ gained along each happiness function $S_i$ from further adding the point $p_j$ to recalculate $\Delta_j$ to use in the next step.

    \begin{algorithm}
    \caption{Multiplicative $1-\frac{1}{e}$-Approximation for $ahr^*$}\label{Alg:AHMS}
    \begin{algorithmic}[1]
    \Require Dataset $D$ with $n$ points, size of ARMS $r$ 
    \Ensure $R$, a $1-\frac{1}{e}$ approximation for $ahr^*$.
    \For {$j \gets 1 $ to $n$}
    \State  $H^{max}_j \gets \max_{p \in D} S_j (p) $ \Comment{so $happ(p,S_j)$ can be computed in $O(d)$}
    \State  $\Delta_j \gets \frac{1}{N}\sum_{i} happ(p_j,S_i)$
    \EndFor
    \State $R_0 \gets \varnothing$
    \For {$j \gets 1 $ to $N$}
    \State {$H_j \gets 0$}
    \EndFor
    
    \For {$s \gets 1 $ to $r$}
    \State Find the point $p_x$ with highest $\Delta_x$
    \State $R_s \gets R_{s-1} \cup \{p_x\}$
    \For {$i \gets 1 $ to $N$}
    \State {$H_i \gets \max(H_i,happ(p_x,S_i))$ }
    \For {$j \gets 1 $ to $n$}
    \State {$\delta_{j,i} \gets \max(happ(p_j,S_i)-H_i,0)$ }
    
    \EndFor
    \EndFor
    \For {$j \gets 1 $ to $n$}
    \State {$\Delta_j \gets \frac{1}{N}(\sum_i\delta_{j,i}) $ }
    \EndFor
    \EndFor
    \State Select $R_r$ as $R$.
    
    \end{algorithmic}
    \end{algorithm}

\begin{theorem}
The given approximation algorithm results in a multiplicative (1-$\frac{1}{e}$) factor approximation and takes $O(drNn)$ time.
\end{theorem}
\begin{proof}
The correctness can be shown through induction. At step $0$, $R_0$, $ahr^*(R_0,S)$ is trivially correct. $\Delta_k$ is recalculated at each step $i$ according to the definitions, and the correctness of $ahr^*(R_i,S)$ follows. The $1-\frac{1}{e}$ approximation bound follows from as discussed as this algorithm follows the greedy procedure of selecting the point with the greatest increase. 

The total time complexity of lines 1-7 is $O(dNn)$ since it computes the value of each function for each point. Line 8-15, which are run $r$ times, take $O(dNn)$ time per iteration, the bottleneck being recomputing the gained happiness contribution for each point from each utility function. In total, the algorithm takes $O(drNn)$ time.
\end{proof}

\subsection{Approximating Average Happiness Ratio for Linear Utilities in 2 Dimensions}
Besides the sampling method, \cite{Zeighami:2016:MAR:2882903.2914831} also proposed an exact dynamic programming algorithm for the special case of linear utilities in $d=2$. As noted in \cite{Zeighami:2016:MAR:2882903.2914831}, this special case is of practical interest as two dimensional datasets often show up after feature selection or extraction. Improving the $O(n^4)$ algorithm from \cite{Zeighami:2016:MAR:2882903.2914831}, we introduce an exact algorithm running in $O(n^2)$ time. As in \cite{Zeighami:2016:MAR:2882903.2914831}, we assume each integral takes constant time to compute. We also present an approximation algorithm running in $O(\frac{n}{\epsilon})$ where $\epsilon$ is the desired additive approximation ratio.

\newcommand{\wa}[0]{\w_\alpha}

\newcommand{\ea}[0]{\eta_\alpha}

For this special case, we have
$ahr(R)= \int_{0}^1 \frac{\max_{p\in R} p \cdot \wa}{\max_{p\in D} p \cdot \wa}\eta_\alpha d \alpha$, where  $\wa$ is the weight vector $[ \alpha, 1-\alpha]$ and $\ea$ is the probability density of $\wa$. Note that this is an equivalent simplification of the definition in \cite{Zeighami:2016:MAR:2882903.2914831}, but differs from the definition in \cite{Storandt2019}, which optimizes $ \frac{\int_{0}^1 (\max_{p\in R} p \cdot \wa)  d \alpha}{\int_{0}^1 (\max_{p\in D} p \cdot \wa) d \alpha}$. 

\begin{figure}
    \caption{Illustration of Dualization}
    \begin{subfigure}[b]{0.2\textwidth}
         \centering
         \includegraphics[width=\textwidth]{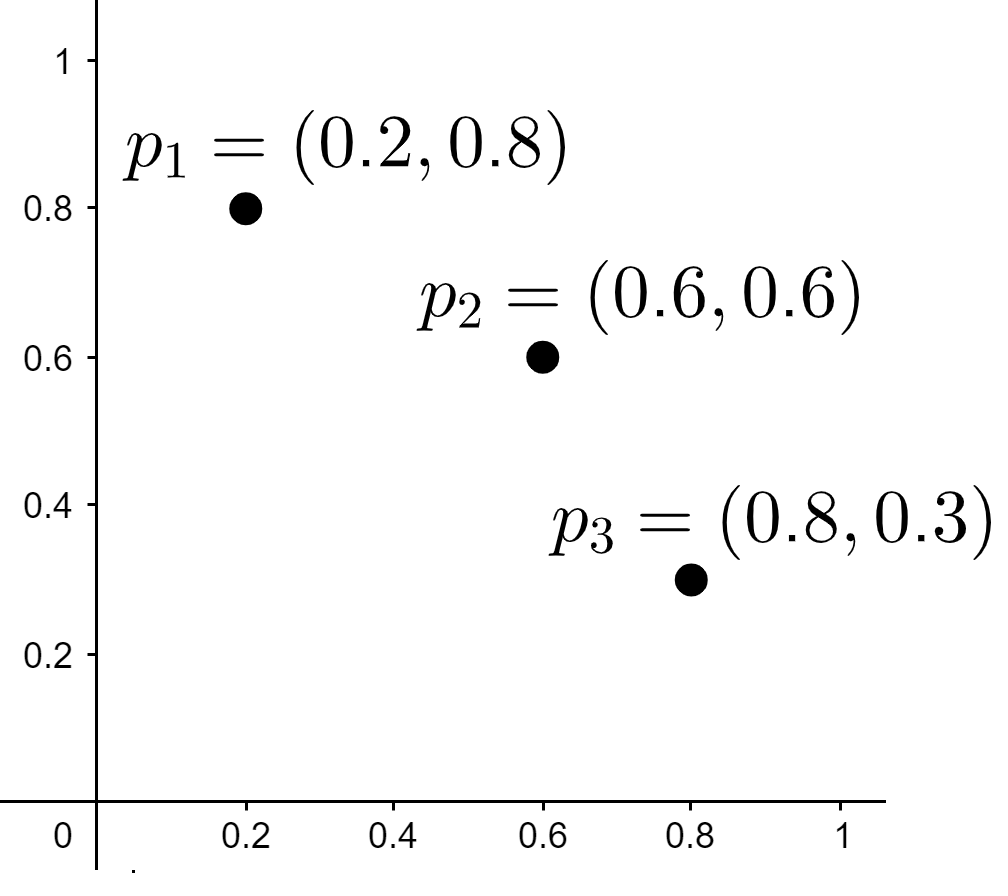}
         \subcaption{Example Dataset $D$}
         \label{fig:y equals x}
     \end{subfigure}
     \hfill
     \begin{subfigure}[b]{0.25\textwidth}
         \centering
         \includegraphics[width=\textwidth]{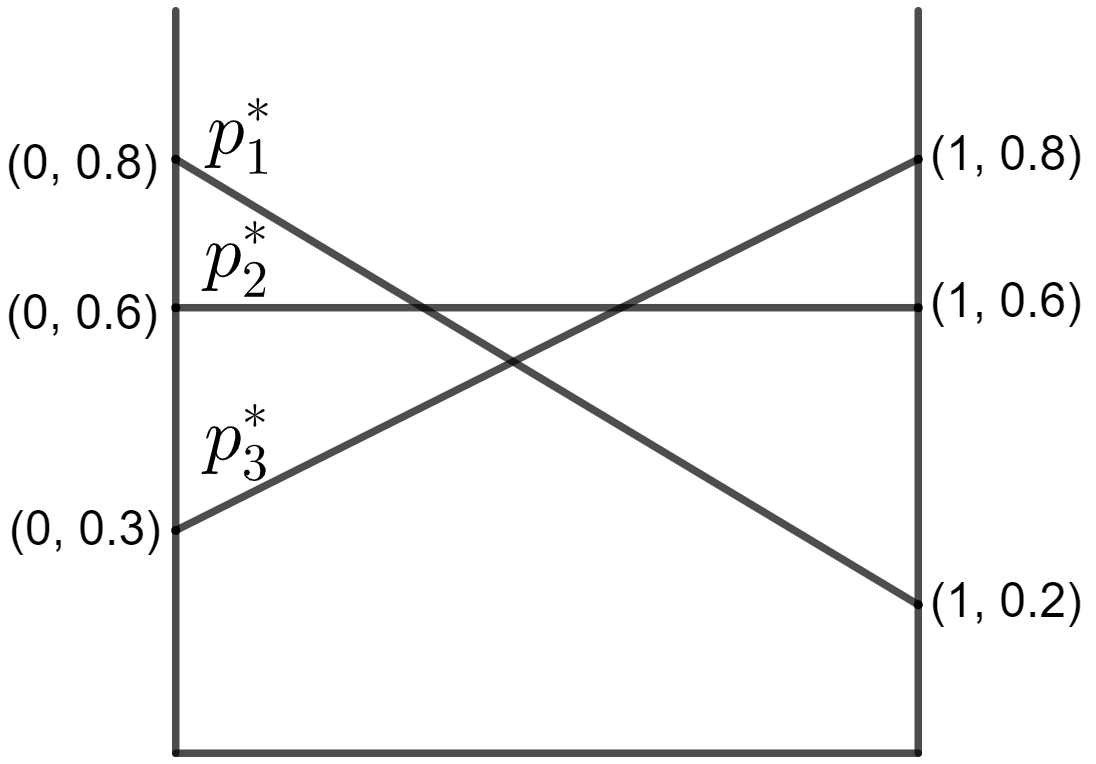}
         \subcaption{Dualization of $D$}
         \label{fig:three sin x}
     \end{subfigure}
     \hfill
    \label{Fig:Dual}
\end{figure}

To aid the proofs, we apply dualization as in \cite{Storandt2019}. Each point $p=(x,y)$ is mapped to its dual line  $p^*$ that passes $(0, y)$ and $(1,x)$ as illustrated in Fig. \ref{Fig:Dual}. It is straightforward to verify that the height of $p^*$ at $x=\alpha$ is equal to $p \cdot \wa$.  Then $\max_{p\in R} p \cdot \wa$ will equal the $y$-value of highest line corresponding to a point in $R$ at $x=\alpha$. It then follows that the contribution to $ahr$ of $\wa$ corresponds the height of the upper envelope of duals of points selected in $R$ at $x=\alpha$ scaled by $ \frac{\eta_\alpha}{\max_{p\in D} p \cdot \wa} $.

\setcounter{figure}{1}

\begin{figure}
    \caption{Illustration for Lemma 5.1}
    \centering
    \includegraphics[width=0.4\textwidth]{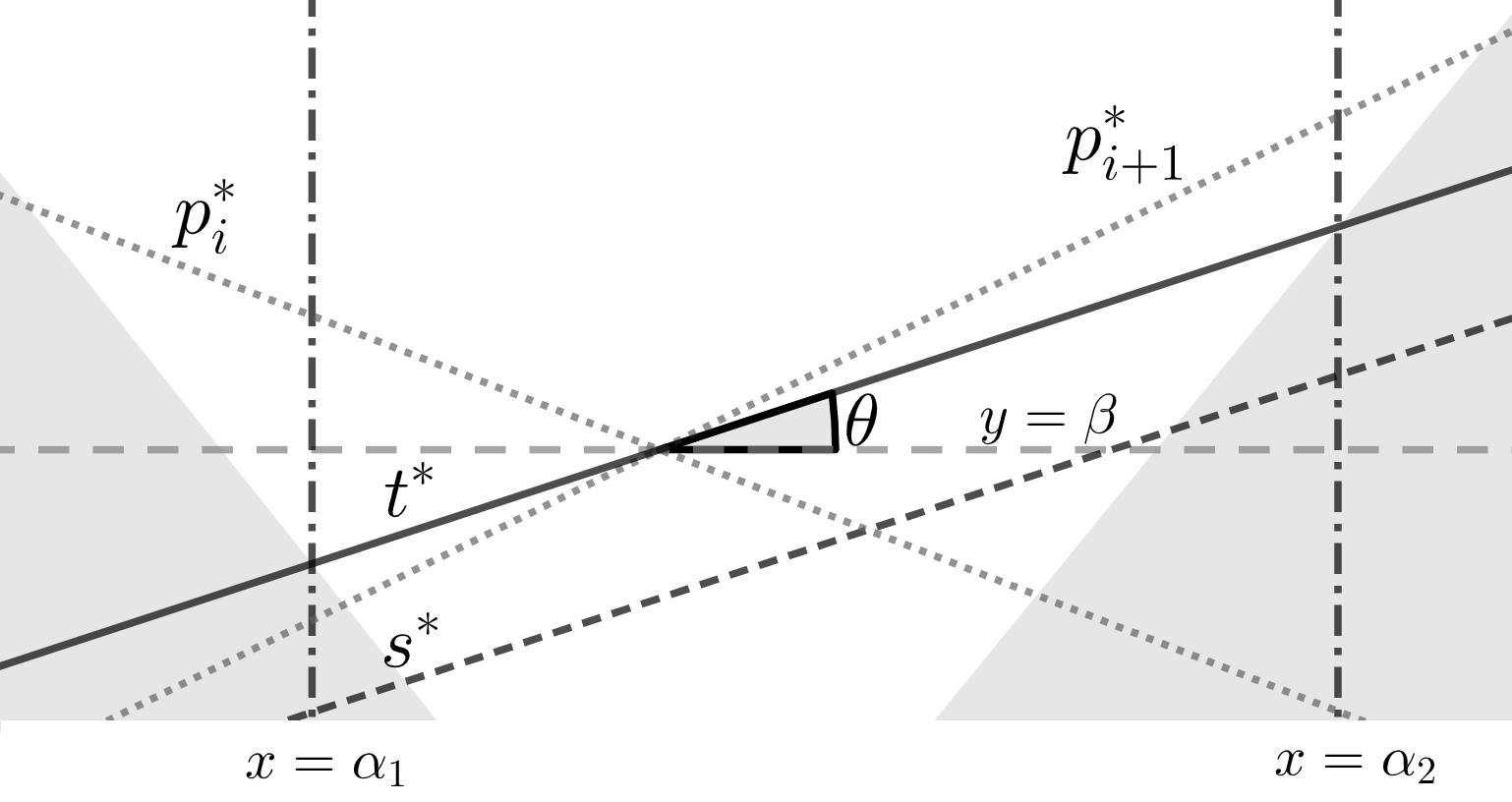}
    \label{Fig:NoNonConvex}
\end{figure}

\begin{figure}
    \caption{Illustration of H[i][j]}
    \centering
    \includegraphics[width=0.4\textwidth]{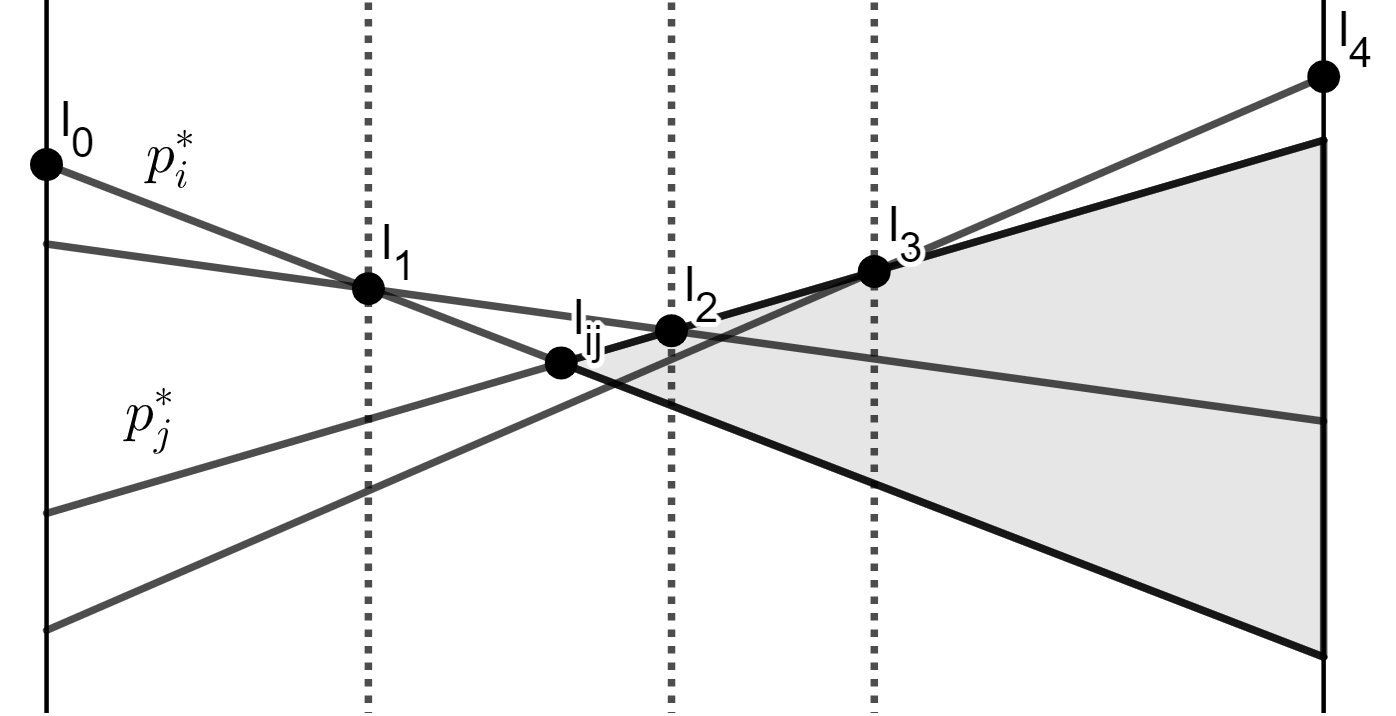}
    \label{Fig:HCalc}
\end{figure}

As we are working with linear utilities, we assume for simplicity that $D$ consists only of skyline points. We now make the following observation: 
\begin{theorem}
There is an optimal ARMS for linear utilities in $d=2$ that consists only of points on the convex hull of $D$.
\end{theorem}

\begin{proof}
First note that any non-convex hull point $s$ must be in the triangle spanned by two adjacent points $p_i, p_{i+1}$ on the convex hull and the origin. We will show it is always optimal to replace $s$ with either $p_i$ or $ p_{i+1}$. 

Assume the optimal set is $R$ and includes $s$. Let $R' = R - \{ s \}$. First note that there is a point $t$ with dual $t^*$ that has the same slope $\tan \theta$ as the dual of $s$, $s^*$ which passes the intersection $(\gamma, \beta)$ between $p_i^*$ and $ p_{i+1}^*$ (as $s$ is not a convex hull point, it is below the intersection and so $t$ dominates $s$). We will show that adding either $p_i$ or $ p_{i+1}$ to $R'$ is no worse than adding $t$, which is no worse than adding $s$. 

If adding $t$ to $R'$ does not add to the $ahr$, then we are done as adding $p_i$ or $p_{i+1}$ cannot make $ahr$ worse. Otherwise, there is a set of values of $\alpha$ where $t$ is better than all points in $R'$ for $\wa$. This set must be a contiguous interval ($\alpha_1$, $\alpha_2$) since it is the intersection of the intervals where $t \cdot \wa > p \cdot \wa$ for each $p \in R'$. This is depicted in Fig. \ref{Fig:NoNonConvex} with shaded regions representing parts already covered by $R'$.  Now, the increase to $ahr$ from adding $t$ to $R'$ is $ \Delta = \int_{\alpha_1}^{\alpha_2} (\tan \theta (\alpha - \gamma) + \beta - m_\alpha) c_\alpha d\alpha$ where $c_\alpha =  \frac{\eta_\alpha}{\max_{p\in D} p \cdot \wa}$ and $m_\alpha = {\max_{p\in R'} p \cdot \wa}$. 

As $t$ is under the section of the convex hull between $p_i$ and $ p_{i+1}$, $\theta \in [\theta_i, \theta_{i+1}]$ where $\theta_i$ and $\theta_{i+1}$ are the slopes of $p_i^*$ and $ p_{i+1}^*$ respectively. The maximum value of $\Delta$ must occur at a local maxima or at one of the endpoints. If it occurs at one of the endpoints, then we have that adding $p_i$ or $p_{i+1}$ is not worse than $t$ as desired. Thus, we need only show that $\Delta$ has no local maxima. 

Using Leibniz's Integral Rule, $ \frac{d \Delta}{d \theta}= \sec^2 \theta \int_{\alpha_1}^{\alpha_2} (\alpha - \gamma) c_\alpha d\alpha - (\tan \theta (\alpha_1 - \gamma) + \beta - m_{\alpha_1}) c_{\alpha_1} \frac{d \alpha_1}{d \theta} + (\tan \theta (\alpha_2 - \gamma) + \beta - m_{\alpha_2}) c_{\alpha_2} \frac{d \alpha_2}{d \theta}  $. By definition of $\alpha_1$ and $\alpha_2$, both $\tan \theta(\alpha_1 - \gamma) + \beta - m_{\alpha_1}$ and $ \tan \theta(\alpha_2 - \gamma) + \beta - m_{\alpha_2}$ are $0$, so $ \frac{d \Delta}{d \theta}= \sec^2 \theta \int_{\alpha_1}^{\alpha_2}  (\alpha - \gamma) c_\alpha d\alpha$. Now, assume there is a local maxima at $\theta =\theta^*$. We obtain $ \frac{d \Delta}{d \theta}\Bigr|_{\theta = \theta^*}= \sec^2 \theta^* \int_{\alpha_1}^{\alpha_2}  (\alpha - \gamma) c_\alpha d\alpha = 0 \implies \int_{\alpha_1}^{\alpha_2}  (\alpha - \gamma) c_\alpha d\alpha = 0$ (as $\sec^2 \theta^* \geq 1$). However, this would imply $ \frac{d \Delta}{d \theta}= 0$ for all values of $\theta$ implying that $\Delta$ is a constant function, contradicting the assumption that there is a local maxima as needed. 
\end{proof}

A consequence of this theorem is that only the convex hull points need to be considered. This leads to a straightforward $O(rn^2)$ dynamic programming algorithm. Let $c$ be the number of points on the convex hull. Labelling the points of the convex hull as $p_1,p_2, \ldots, p_c$ in order of increasing x value (which implies increasing slope), and adding $(0,0)$ as $p_0$ for simplification, we have the recurrence relation $D[k][j] = \max_{0 \leq i<j} D[k-1][i] + H[i][j]$, where $H[i][j]$ is the increase in happiness from adding point $p_j$ to the set after most recently adding $p_i$ as illustrated in Fig. \ref{Fig:HCalc} --- any points added before $p_i$ do not affect the happiness increase from adding $p_j$, since $p_i^*$ will intersect $p_j^*$ at a larger x coordinate than it does $p_{i'}^*$ for $i'<i$ --- this is implied by Lemma 5.1. 

\begin{lemma}
If $0 \leq i < j < k \leq c$, $I_{ik} \leq I_{jk} $ where $I_{ab}$ is the x coordinate of the intersection between $p_a^*$ and $p_b^*$. 
\end{lemma}
\begin{proof}

First notice that for $a<b$, $\alpha < I_{ab} \implies$ $\wa \cdot p_a > \wa \cdot p_b$. Similarly, $\alpha > I_{ab} \implies$  $\wa \cdot p_a < \wa \cdot p_b$ for and $\alpha = I_{ab} \implies$ $\wa \cdot p_a = \wa \cdot p_b$. 

Assume $I_{ik} > I_{jk}$ for the sake of contradiction. Then for $\alpha > I_{ik}$, we have $p_k \cdot \wa > p_j \cdot \wa$ since $\alpha > I_{ik} \implies \alpha > I_{jk}$. At $\alpha = I_{ik}$ , since $I_{ik} > I_{jk}$, we have  $p_i \cdot \wa  = p_k \cdot \wa > p_j \cdot \wa$. For $\alpha < I_{ik}$, we must have $p_i \cdot \wa > p_j \cdot \wa$ since $p_i \cdot \wa > p_j \cdot \wa$ at $\alpha=I_{ik}$ and the difference only increases as $\alpha$ decreases because the slope of $p_i^*$ is less than that of $p_j^*$. Thus, for any utility $\wa$, at least one of $p_i$ or $p_k$ will be better than $p_j$ for $\wa$, contradicting the fact that $p_j$ is a convex hull point (a convex hull point must be optimal for at least one utility function). 
\end{proof}

We first show $H[i][j]$ can be filled in $O(n^2)$. It is helpful to refer to Fig. \ref{Fig:HCalc}. Here, we label the x coordinates of the intersections of the duals on the upper envelope $I_1, I_2, \dots, I_{c-1}$. We set $I_0 = 0$ and $I_{c} = 1$ for simplicity. It can be seen that for a pair of points $p_i$, $p_j$, the added happiness (corresponding to shaded area) can be divided by $x$ coordinate into two parts: A) the section between the intersection of $p_i^*$ and $p_j^*$ and the next intersection on the upper envelope ($I_{ij}$ to $I_2$ in the figure) and B) sections completely spanning a segment of the upper envelope (($I_2$, $I_3$) and ($I_3$, $I_4$) in the figure). Notice that the point $p_i$ optimizes $\wa$ for $\alpha \in [I_{i-1},I_i]$. Thus, the contribution to $ahr$ of a point $s$ in $[I_{a},I_b] \subset [I_{i-1},I_i]$ is simply $F(q, p_i, I_{a}, I_{b}) = \int_{I_{a}}^{I_{b}} \frac{q \cdot \wa}{p_i \cdot \wa} \ea d \alpha$. This has a closed form which can be computed in $O(1)$ if $\ea = 1$ (if $\alpha$ follows a uniform distribution). Then, part A of the added happiness is $F(p_i, p_k, I_{ij}, I_{k}) - F(p_j, p_k, I_{ij}, I_{k})$, where $I_k$ is the intersection immediately after $I_{ij}$, while part B is $\Sigma_{l=k+1}^{c} (F(p_i, p_l, I_{l-1}, I_{l}) - F(p_j, p_l, I_{l-1}, I_{l}))$  --- this is the happiness of $p_i$ minus that of $p_j$ in each region. Part B can be calculated efficiently with prefix sums $S[i][k] = \Sigma_{l=i}^{k} F(p_i, p_l, I_{l-1}, I_{l})$. This is expressed as an algorithm in Algorithm \ref{Alg:HCalc}, where it can be seen that $F$ is computed $O(n^2)$ times, each representing one integral computation, while other operations also take $O(n^2)$ time.

    \begin{algorithm}
    \caption{Computing $H$}\label{Alg:HCalc}
    \begin{algorithmic}[1]
    \Require Convex hull points $p_1,p_2, \ldots, p_c$ 
    \Ensure $H[i][j]$ for $j \geq i$
    \State Compute the x coordinates of the intersections on the upper envelope as $I_1, I_2, \dots, I_c$. 
    \For{$i \gets 1$ to $c$} 
    \State $S[i][i]  \gets F(p_i, p_i, I_{i-1}, I_i)$
    \For{$k \gets i$ to $c$}   
    \State $S[i][k] \gets  F(p_i, p_j, I_{k-1}, I_k) + S[i][k-1]$ 
    \EndFor
    \EndFor
    \For{$i \gets 1$ to $c$} 
    \State $k \gets 0$
    \For{$j \gets i+1$ to $c$}
    \State Compute the x coordinate of the intersection between $p_i^*$ and $p_j^*$ as $I_{ij}$ (treating $p_0^*$ as the y axis to simplify).
    \While {$I_{ij} \geq I_{k+1}$} \Comment{ $I_{ij}$ is increasing in $j$ (Lemma 5.1)}
    \State $k \gets k+1$
    \EndWhile
    \State $A \gets  F(p_i, p_k, I_{ij}, I_k) -  F(p_j, p_k, I_{ij}, I_k)$
    \State $B \gets  (S[i][c]-S[i][k]) - (S[j][c]-S[j][k]) $
    \State $H[i][j] \gets A + B$
    \EndFor
    \EndFor
    \end{algorithmic}
    \end{algorithm}

Now that we have shown $H[i][j]$ can be filled in $O(n^2)$ integral computations, the given recurrence already results in a $O(rn^2)$ algorithm. However, this can further be improved by noticing that $H[i][j]$ satisfies a \textbf{Quadrangle Inequality}, also called the \textbf{Inverse Monge} property \cite{MongeProp}: 

\begin{figure}
    \caption{Illustration of Quadrangle Inequality}
    \centering
    \includegraphics[width=0.4\textwidth]{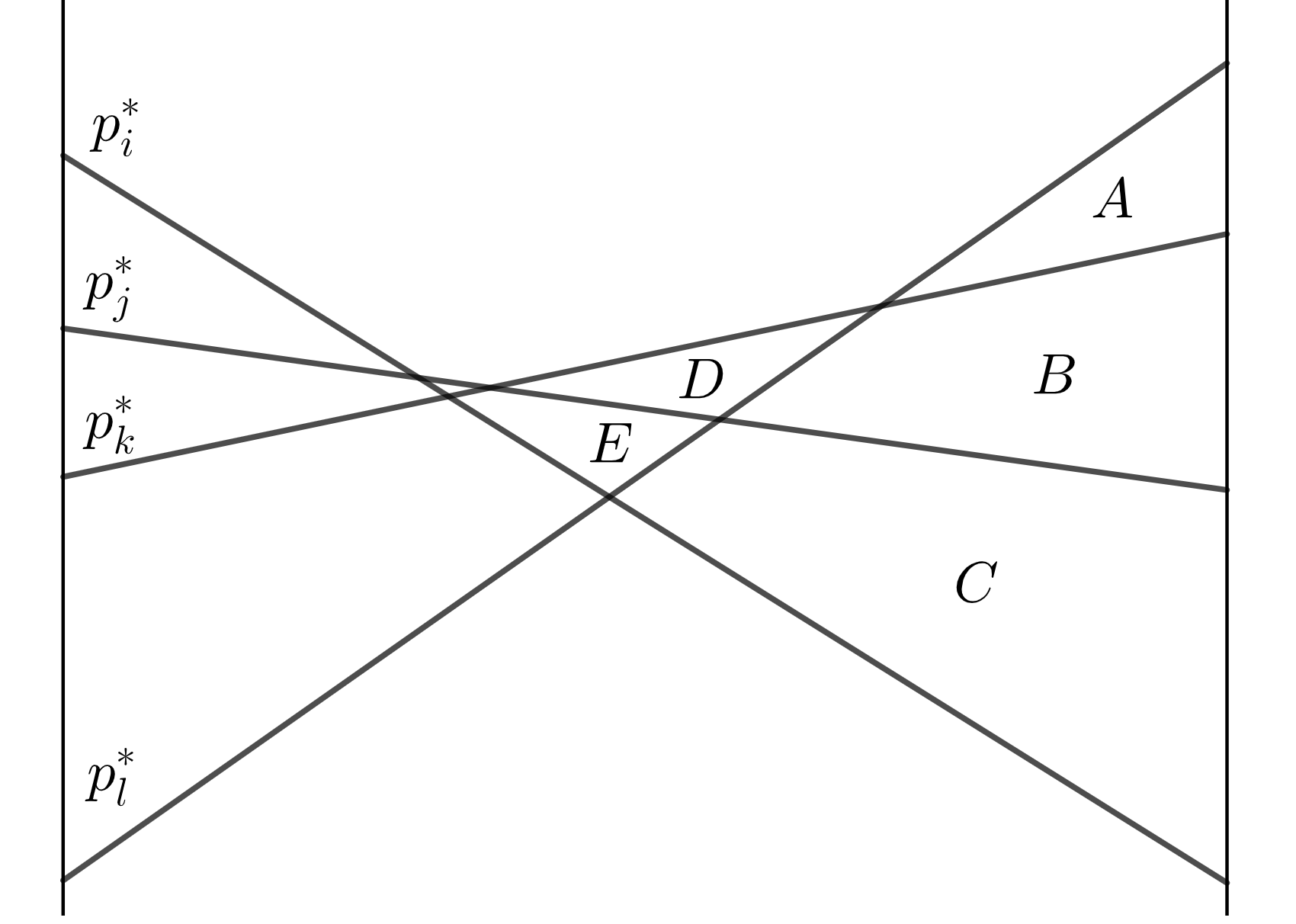}
    \label{Fig:Quad}
\end{figure}

\begin{lemma}
For $0 \leq i< j < k < l \leq c$ , $ H[i][l] + H[j][k] \leq H[i][k] + H[j][l]$
\end{lemma}
\begin{proof}
As shaded in Fig. \ref{Fig:HCalc}, $H[i][j]$ is the happiness from covering the area between $p_i^*$ and $p_j^*$ to the right of their intersection. To simplify the proof, we treat the dual of $p_0=(0,0)$,  $p_0^*$ as the y axis. 

Let $A,B,C,D,E$ be the added happiness from covering each area as labelled in  Fig. \ref{Fig:Quad} where $i < j < k < l$. We have $H[i][l] = A +  B + C$, $H[j][k] = B+D $, $H[i][k] =  B + C + D + E$ and $H[j][l] = A + B$. Then, $ (H[i][k] + H[j][l]) - (H[i][l] + H[j][k]) = E \geq 0$ $\implies  H[i][l] + H[j][k] \leq H[i][k] + H[j][l]$.
\end{proof}

As shown in \cite{Monge}, this property allows us to fill out recurrences of the form  $D[k][j] = \max_{0 \leq i<j} D[k-1][i] + H[i][j]$ for $ 1 \leq k \leq r, 1 \leq j \leq c$ in $O(rn)$. Essentially, the main idea is that this inverse Monge property allows for the application of the SMAWK algorithm \cite{smawk} to compute $D[k][j], 1 \leq j \leq c$ for each fixed $k$ in $O(n)$ time. This follows from the following facts:

\begin{itemize}
    \item If $H$ fulfills the inequality in Lemma 5.2, then so does $H'_{k-1}$ where $H'_{k-1}[i][j] = D[k-1][i] + H[i][j]$ for $i< j$ ($ H[i][l] + H[j][k] \leq H[i][k] + H[j][l]\implies  D[k-1][i] + H[i][l]  +  D[k-1][j] +H[j][k] \leq  D[k-1][i] + H[i][k] + D[k-1][j] + H[j][l]$). As mentioned, in \cite{Monge}, only the elements above the diagonal are relevant. Nevertheless, we can set $H'_{k-1}[j][i] = j - i $ for $j<i$ to obtain a full inverse Monge matrix to simplify the proof without affecting the algorithm's correctness.  
    \item  An \textbf{Inverse Monge} matrix is \textbf{Totally Monotone}, meaning that for each submatrix, the row indices of the maximum value in each column (taking the last row in case of ties) are non-decreasing. \cite{MongeProp}
    \item The SMAWK algorithm can find the column maximums of a $n$ by $n$ \textbf{Totally Monotone} matrix in $O(n)$ \cite{smawk}. 
\end{itemize}

Combining these ideas, we obtain Algorithm \ref{Alg:2DAHMS}. Finding the skyline and the convex hull points takes $O(n \log n)$. Computing $H$ takes $O(n^2)$. SMAWK is run $O(r)$ times, each taking $O(n)$ time. Thus, the overall runtime is $O(n^2)$. Note that as the same optimal solution optimizes both regret and happiness ratios, this exact algorithm optimizes both the regret and happiness. 

    \begin{algorithm}
    \caption{Exact Algorithm for 2D-ARMS with Linear Utilities}\label{Alg:2DAHMS}
    \begin{algorithmic}[1]
    \Require Dataset $D$ with $n$ points, size of ARMS $r$
    \Ensure $R$, the ARMS
    \State Find the convex hull of the skyline points $p_0, p_1, \dots, p_c$ sorted by x coordinate.
    \State Compute $H$ as in Algorithm \ref{Alg:HCalc}
    \For{$i \gets 1$ to $c$}   
    \State $D[1][i] \gets  H[0][1]$  \Comment{$H[0][i]$ is equivalent to $ahr({p_i})$}
    \EndFor
    \For{$k \gets 2$ to $r$} 
    \State Apply SMAWK to compute $D[k][i]$ for $1 \leq i \leq c$ from $H'_{k-1}$ as defined above. ($H'_{k-1}$ is not explicitly constructed, but its entries can be calculated on the fly as needed in SMAWK)
    \EndFor
    \State Find the $i$ that maximizes $D[r][i]$ and select the points that constructed it as $R$. 
    \end{algorithmic}
    \end{algorithm}

    \begin{algorithm}
    \caption{Additive $\epsilon$-Approximation for 2D-ARMS with Linear Utilities}\label{Alg:Approx}
    \begin{algorithmic}[1]
    \Require Dataset $D$ with $n$ points, size of ARMS $r$, Additive approximation ratio $\epsilon$
    \Ensure $R$, an $\epsilon$-approximate ARMS
    \State Find the convex hull of the skyline points $p_0, p_1, \dots, p_c$ sorted by x coordinate.
    \State Apply the \textbf{Additive Reduction Scheme} with factor $\epsilon$ on the convex hull and again find the convex hull of the reduced points to obtain $c'$ candidate points. 
    \State Compute $H$ with a modified version of Algorithm \ref{Alg:HCalc}, passing both the convex hull and the candidate points.
    \State Construct $R$ from $H$ as in Algorithm \ref{Alg:2DAHMS}.

    \end{algorithmic}
    \end{algorithm}
    
Finally, we note Algorithm \ref{Alg:2DAHMS} can be turned to an approximation algorithm, Algorithm \ref{Alg:Approx}, with additive approximation ratio $\epsilon$ running in $O(\frac{n}{\epsilon} + n \log n)$ by applying the \textbf{Additive Reduction Scheme} (4.1.1) to obtain a list of at most $O(\frac{1}{\epsilon})$ candidate points (not $O(\frac{1}{\epsilon^2})$ as we only consider the skyline). The main modification is to the computation of $H$, where we must now also pass a list of candidate points along with the convex hull points --- $H$ becomes a $c' \times c'$ matrix where $c'$ is the number of candidates. Specifically, we change the loops in lines 2, 6, and 8 of Algorithm \ref{Alg:HCalc} to loop through the candidate points instead of the whole convex hull. This changes the time complexity of computing $H$ to $O(\frac{n}{\epsilon})$. 

    This algorithm finds the set $R$ only containing candidate points with the maximum happiness (the proof follows near identically to the exact case, only changing the considered points in the calculation of $H$ and $D$). Thus, the approximation ratio can be proven as in Lemma 4.3 only replacing the max operators in the numerator with the expectation. Filling out $D$ now takes only $O(\frac{r}{\epsilon})$, so the overall runtime becomes $O(\frac{n}{\epsilon} + n \log n)$. As in Section 4, this approximation ratio for the \textbf{Additive Reduction Scheme} applies to both regret and happiness. 

\section{Experimental Results}
This section presents experimental results for our reduction schemes and our approximation algorithms for ARMS. The algorithms were implemented in C++ and run on Ubuntu 16.04 VirtualBox virtual machine with 12GB RAM and 6 2.5 GHz Intel Core i7 Processors. The scalability tests in particular are run on CentOS IBM X3650 M3 servers with 2x6-Core 2.66GHz and 48GB RAM.

\subsection{Polynomial Time Reduction Schemes}

In this subsection, we discuss experimental results for applying our reduction schemes as a preprocessing step before applying existing heuristic algorithms for the 1-RMS problem. 

The experiments were performed to evaluate the effect of applying the reduction schemes on the runtime and achieved Minimum Happiness Ratio (MHR) of selected RMS solvers:DMM \cite{Asudeh:2017:ECR:3035918.3035932}, Geogreedy \cite{Peng2014}, Greedy \cite{Nanongkai2010}, and ImpGreedy \cite{Xie2018}. Note that since the MHR is equivalent to the 1 minus the Maximum Regret Ratio (MRR), the solvers also optimize MHR.

\begin{figure}
    \caption{Reduced Dataset Size vs $\varepsilon$ ($n = 1,000,000$, $d = 5$)}
    \centering
    \includegraphics[width=0.4\textwidth]{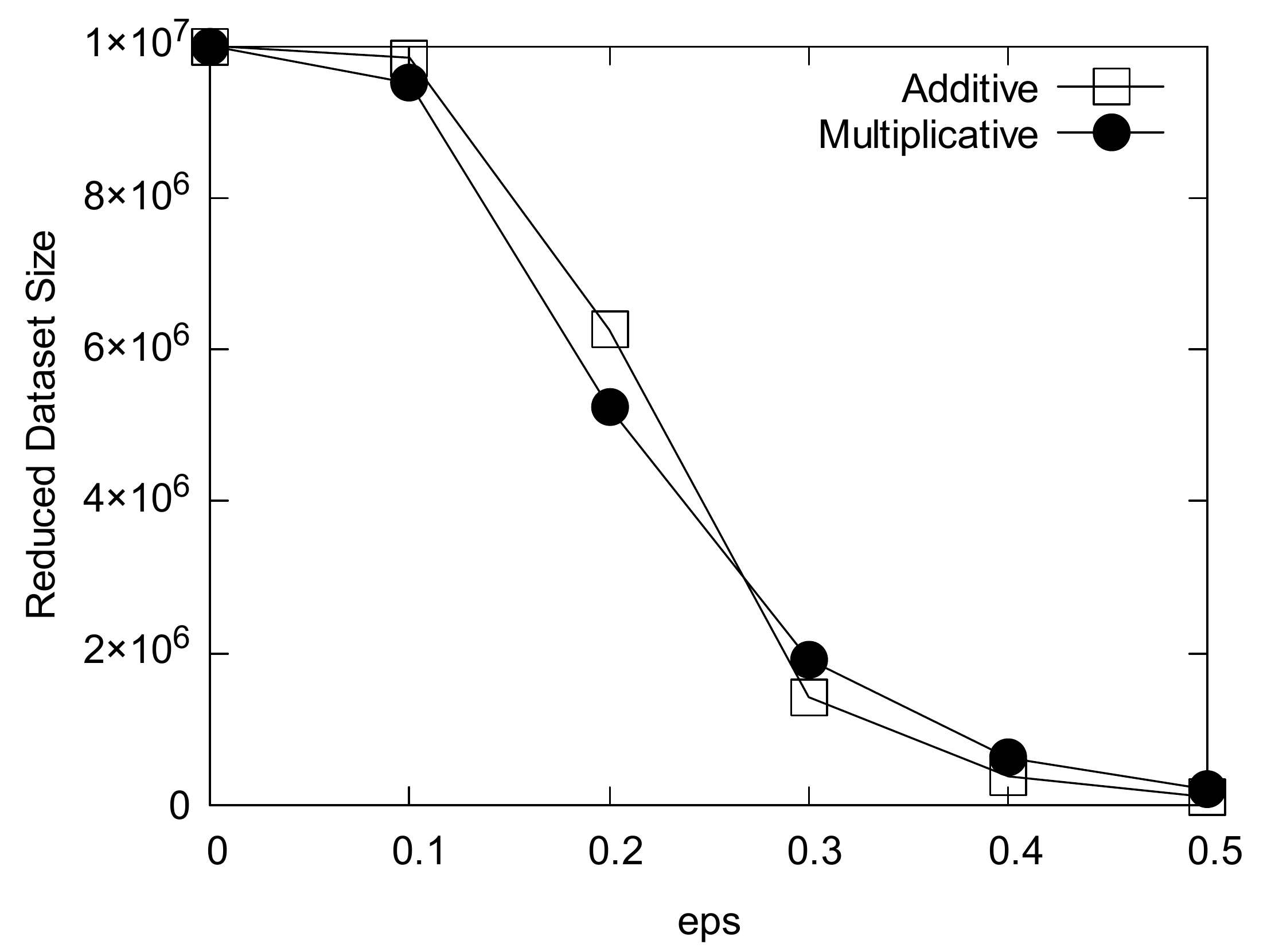}
    \label{Fig:red}
\end{figure}

We first evaluated the reductive power of both schemes at various values of $\varepsilon$ on a 5-dimensional dataset of $10^7$ points as shown in Fig. \ref{Fig:red}. It can be seen that both schemes are effective at reducing the dataset especially for higher values of $\varepsilon$, with the dataset reduced by at least $80\%$ for $\varepsilon > 0.3$. As we consider multiplicative approximation to be the more common goal, we focus on the multiplicative scheme for the remainder of the experiments. As a simple heuristic, when mapping a point in reduced dataset back to the original dataset, we select the original point with the maximum sum of coordinate values. 

\begin{figure}
    \vspace{-5mm}
    \caption{Vary $n$ experiments ($d = 5$, $r = 50$)}
    \vspace{-7mm}
    \centering
    \includegraphics[width=0.4\textwidth]{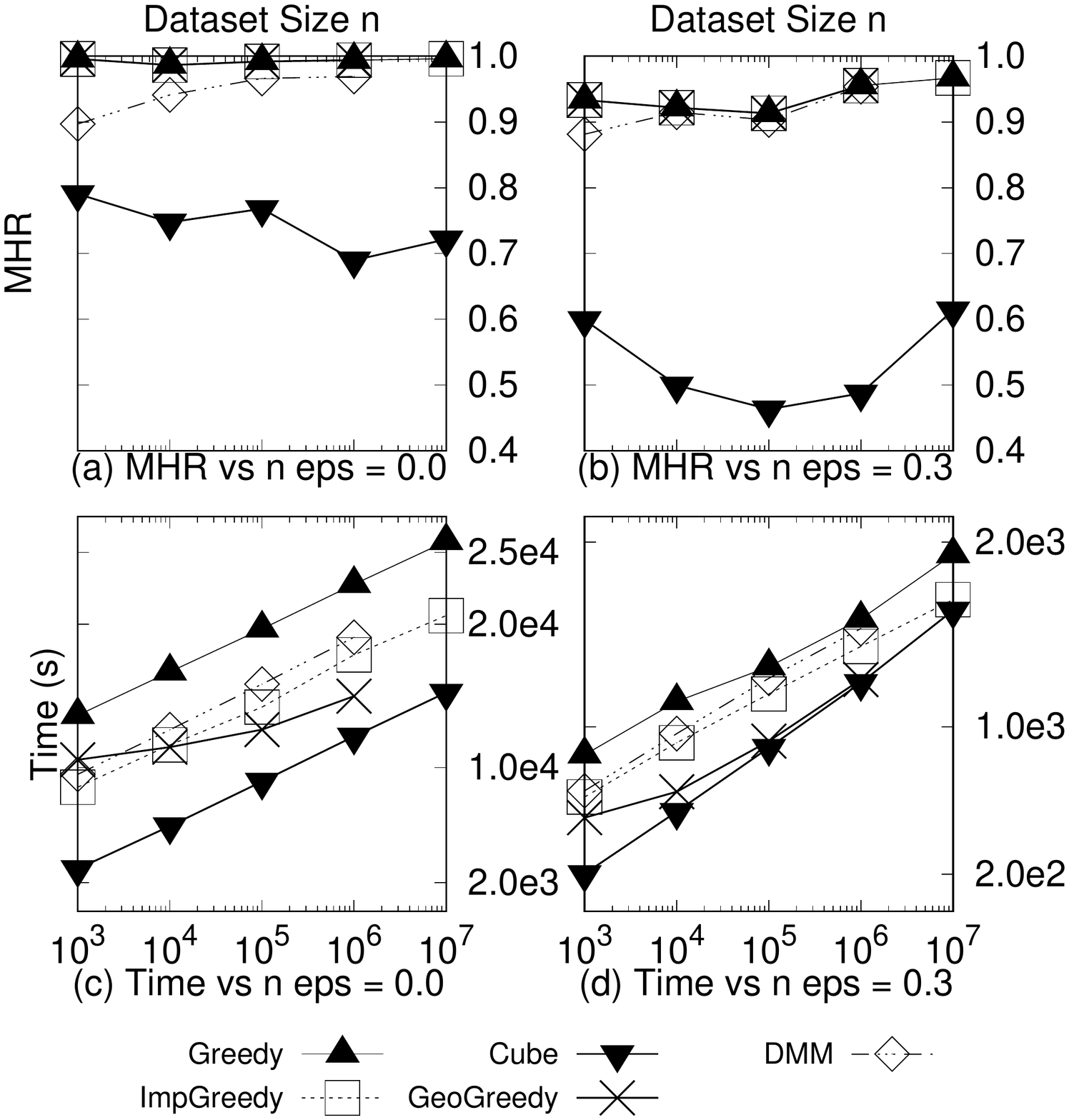}
    \vspace{-18mm}
    \label{Fig:VaryN}
\end{figure}

We conducted a scalability test varying $n$ with and without the multiplicative reduction scheme with $\varepsilon=0.3$ on a randomly generated dataset generated as specified in \cite{Skyline}. Skyline queries were not applied in these experiments (as they should not be in $k$-RMS for $k>1$). These results are shown in Fig. \ref{Fig:VaryN}. 
DMM \cite{Asudeh:2017:ECR:3035918.3035932} and GeoGreedy \cite{Peng2014} have extremely inefficient runtime for large datasets $n \geq 10^7$ and are omitted from the plot in such cases.

As seen in Fig. \ref{Fig:VaryN}, the multiplicative reduction scheme can reduce the required runtime of 1-RMS solvers by up to 92\% (from 27480s to 2138s) while keeping the MHR within 96\% of the original on the largest tested settings. However, for that same setting, the MRR increased by up to a factor of 8.8, which suggests that the reduction schemes are inappropriate for multiplicatively approximating MRR.

\subsection{Approximation Algorithm for ARMS}

\subsubsection{$1-\frac{1}{e}$-approximation algorithm for $ahr^*$}

Here, we experimentally compare the $O(drNn)$ time $1-\frac{1}{e}$-approximation algorithm for $ahr^*$ proposed in Section 3.3.1 with Greedy Shrink FAM \cite{Zeighami:2016:MAR:2882903.2914831}, the $O(dNn^3)$ time algorithm for ARMS, on a 5-dimensional NBA dataset of 17265 points, with 100 sampled linear utility functions. Both were implemented without any heuristics. 

\begin{figure}
\caption{Comparison with Greedy Shrink FAM}
\includegraphics[width = 0.45\textwidth]{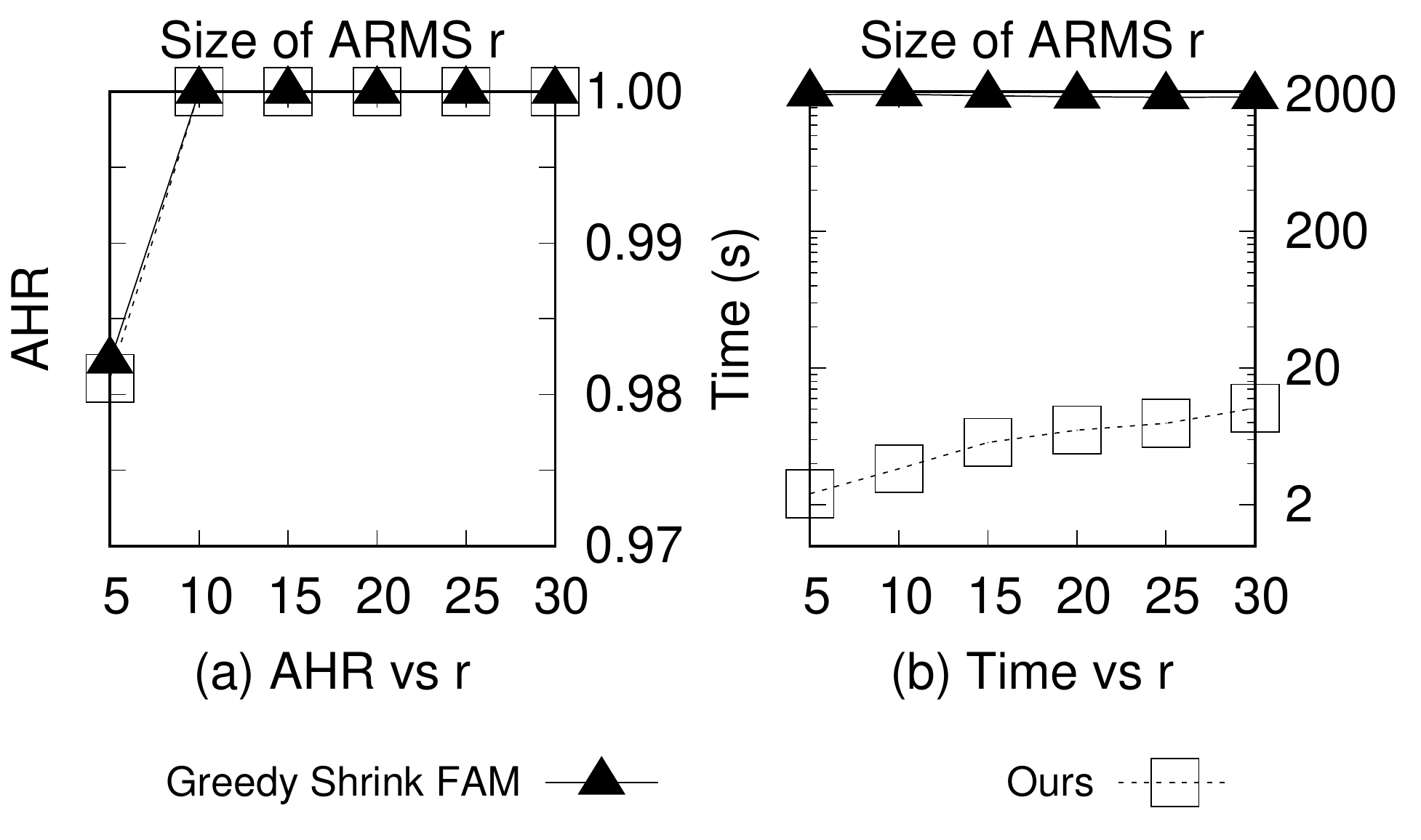}
\label{Fig:AHMSvFAM}
\vspace{-4mm}
\end{figure}

We show our results in Fig. \ref{Fig:AHMSvFAM}. Both algorithms achieve virtually identical average happiness ratios, while ours requires significantly less time to execute, requiring runtime at least two orders of magnitude lower. This result reflects the superior time complexity of our proposed approximation algorithm, despite both having equivalent goals of minimizing regret and maximizing happiness.

\begin{figure}
    \caption{ARMS Vary $n$ experiments ($r = 5$, $d=7$, $N = 1000$)}
    \centering
    \includegraphics[width=0.45\textwidth]{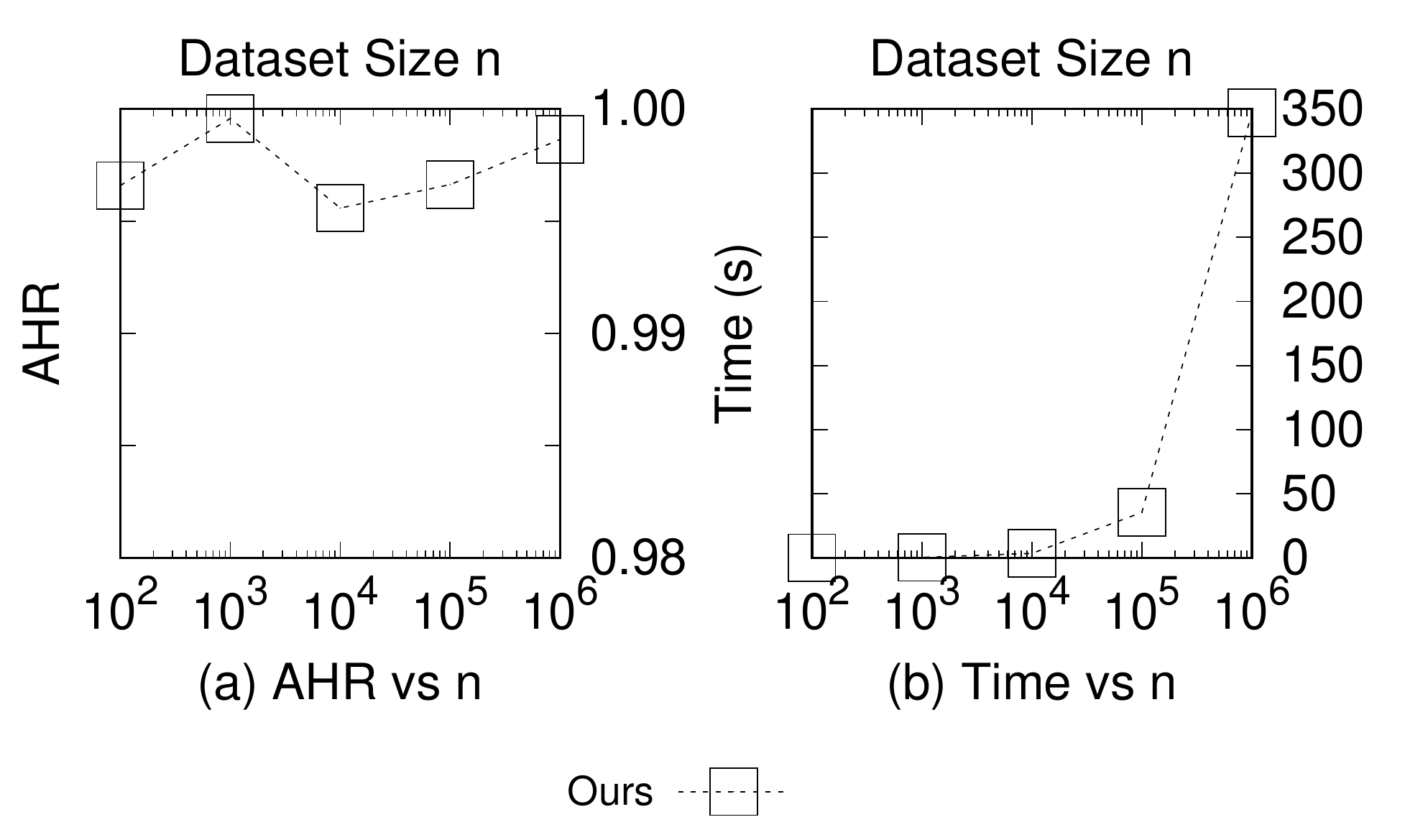}
    \label{Fig:AHMS-VaryN}
\vspace{-4mm}

\end{figure}


We further conducted a scalability test of our algorithm to show the efficiency at various large values of $n$ shown in Fig. \ref{Fig:AHMS-VaryN}. 
The datasets were generated as in \cite{Skyline}. Notably, our algorithm is able to handle datasets of size $10^6$ with more sampled utilities in significantly less time than Greedy Shrink FAM requires for only 17265 points.

\subsubsection{Algorithms for Linear Utilities in $d=2$}

\begin{figure}
    \caption{2D ARMS Vary $n$ experiments ($r = 5$)}
    \centering
    \includegraphics[width=0.45\textwidth]{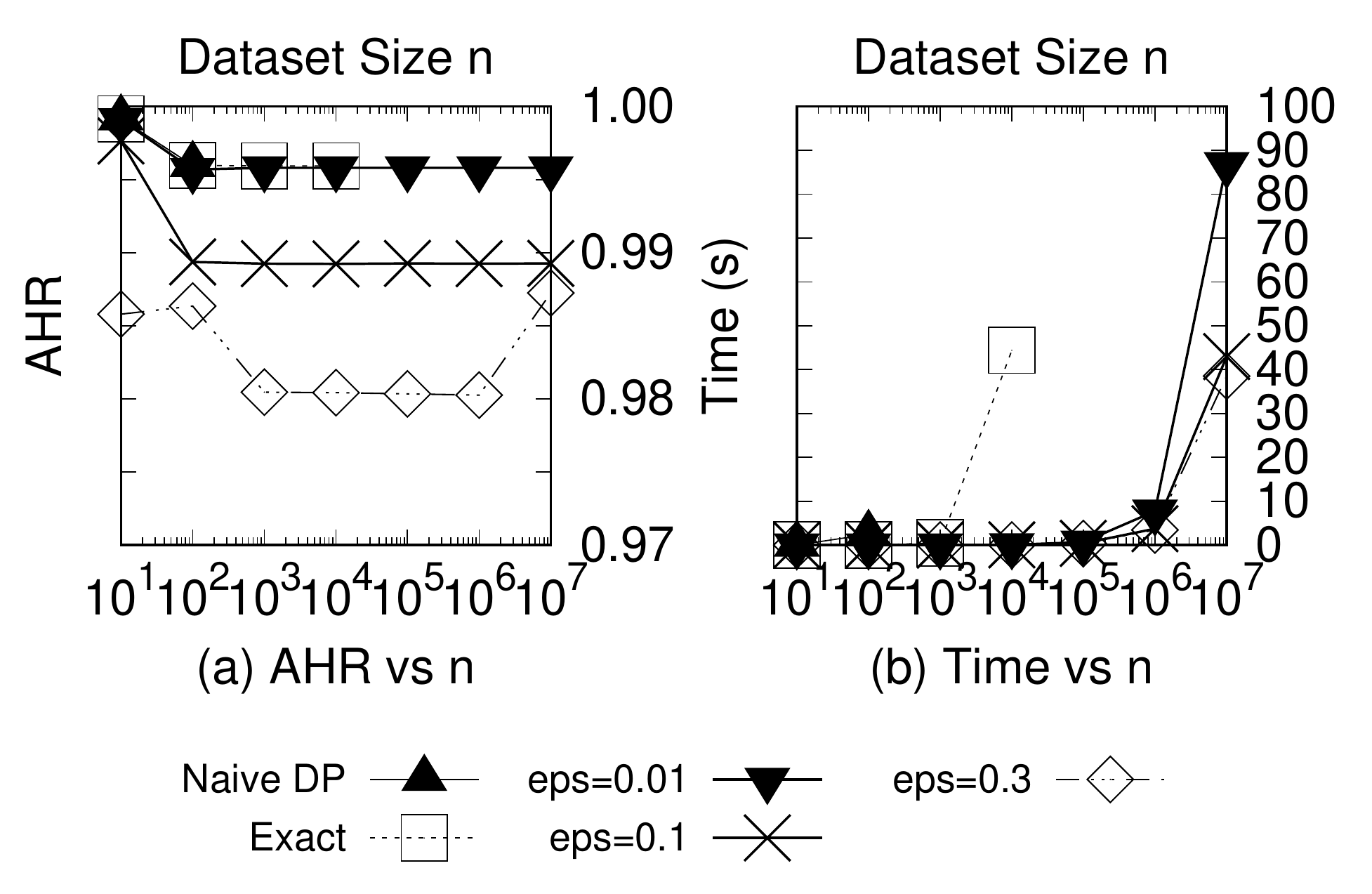}
    \label{Fig:AHMS-2D}
\vspace{-4mm}
\end{figure}

For our $d=2$ algorithm, the time complexity depends significantly on the number convex hull points, however this number can be much smaller than the skyline (for example, it is known that the expected number of vertices on the convex hull of $n$ points uniformly sampled from the unit square is $O(\log n)$ \cite{polytope}). Thus, we chose to generate input data by sampling points on the unit circle in the first quadrant. This guarantees that all points are on the convex hull and lets us show the empirical time complexity in terms of convex hull size. Note that this can only make our algorithms' performance worse without affecting the performance of the naive DP algorithm \cite{Zeighami:2016:MAR:2882903.2914831}, which is agnostic to whether a point is on the convex hull. As for the utility function distribution, we set $\eta_\alpha=1$ for all $\alpha$ (uniform distribution) so that the integrals have a closed form as previously discussed.

We performed an experiment varying $n$ for the DP algorithm \cite{Zeighami:2016:MAR:2882903.2914831}, our Exact algorithm, and our approximation algorithms with $\epsilon = 0.01, 0.1,$ and $  0.3$. This is shown in Fig. \ref{Fig:AHMS-2D}, omitting points for when an algorithm exceeds memory limits. 

It can be seen that our Exact algorithm can handle significantly larger datasets than DP, being successful on $n=10^4$, while DP already exceeds memory limits at $n=10^3$. This reflects their respective memory complexities --- $O(n^3)$ vs. $O(n^2)$. In terms of Average Happiness Ratio (AHR), as both algorithms are exact, they achieve the same AHR when successful. 

Meanwhile, our approximation algorithms handled the largest tested dataset of $10^7$ points, with the slowest requiring only 87 seconds. This shows that the additive reduction is successful at reducing the time and memory needed, while only marginally affecting AHR (AHR $\geq$ 0.995 for $\epsilon=0.01$ on all tested sets). We also note that the time reduction between $\epsilon=0.1$ and $\epsilon=0.3$ is barely noticeable while going from $\epsilon=0.01$ to $\epsilon=0.1$ reduces the time by roughly half. This reflects its time complexity, $O(\frac{n}{\epsilon} + n \log n)$, with the $O(n \log n)$ factor dominating as $\epsilon$ becomes sufficiently large.

\subsection{Experimental Summary}

For $k$-RMS, our multiplicative reduction scheme reduced the runtime of 1-RMS solvers by up to 92\% while keeping the MHR within 96\% of the original on the largest tested settings. 

For the sampling based approach for AHMS, our algorithm ran in less than 1\% of the time used by Greedy Shrink FAM while maintaining nearly identical AHR on the largest settings. For the special case of linear utilities in $d=2$, our exact and approximation algorithms could handle datasets of $10^4$ and $10^7$ points respectively, whereas the existing DP algorithm failed on even $10^3$.

\section{Conclusion and Future Work}

We have studied the approximation of the happiness maximization version of regret minimizing set problems, resulting in multiple algorithms which come with stronger theoretical guarantees or time complexities than existing algorithms.

For $k$-RMS, have completely resolved the NP-Hardness of multiplicatively approximating $k$-happiness for all values of $d$ and $k$ by showing that it is multiplicatively approximable to any desired ratio for fixed $d$, but NP-Hard to multiplicatively approximate for unfixed $d$. 
We then introduced dataset reduction schemes which we experimentally show to significantly improve the runtime of existing heuristic algorithms while mostly preserving the happiness ratio. 
Finally, for ARMS, we have provided approximation ratios with significantly improved time complexities for the average happiness ratio. In particular, our algorithm for optimizing the average happiness of a sample is faster by a factor of $\Theta(n^2/r)$ than the algorithm previously proposed, while our exact algorithm for the special case of linear utilities in 2 dimensions improves on the previous by a factor of $\Theta(n^2)$.

For future work, as the provided polynomial-time approximation schemes for $k$-RMS are intended primarily as theoretical tools, it remains open whether computationally feasible schemes exist. 

\bibliographystyle{ACM-Reference-Format}
\bibliography{bib}

\end{document}